%% file: main.tex
\documentclass[11pt]{article}
\usepackage[margin=1in]{geometry}
\usepackage{fmtcount} % 1st, 2nd...
\usepackage[english]{babel}
\usepackage{soul} % spacing
\usepackage{enumerate} % change enum label
\usepackage{framed} % frame pagebreak
\usepackage{authblk}
\usepackage{setspace}
\usepackage{import}

\usepackage{amsmath}
\usepackage{amsfonts}
\usepackage{amssymb}
\usepackage{amsthm}
\usepackage{amscd}
\usepackage{array}
\usepackage{mathtools} 
\usepackage{verbatim}
\usepackage{fancyvrb} % verbatim
\usepackage{algpseudocode}
\usepackage{algorithm}
\usepackage{upgreek, bm}

\usepackage{graphicx}
\usepackage{epstopdf}
\usepackage{xcolor}
\usepackage{caption}
\usepackage{subcaption}
\usepackage{multirow} % table
\usepackage{longtable} % table to next page
\usepackage{rotating}
\usepackage{wrapfig} % wrap fig, tbl with text

\usepackage{url}
\usepackage{cite}
\renewcommand\Affilfont{\normalsize}

\DeclareMathOperator*{\argmin}{arg\,min}

\DeclareMathOperator*{\bbr}{\mathbb{R}}
\newcommand{\1}{{\rm 1}\kern-0.24em{\rm I}}
\newcommand{\cN}{\mathcal{N}}
\newcommand{\cO}{\mathcal{O}}
\newcommand{\ctN}{\mathcal{N}_\tau}

\title{Multidimensional scaling informed by $F$-statistic: Visualizing grouped microbiome data with inference}

\author[1,2]{Hyungseok Kim$^*$}
\author[3]{Soobin Kim$^*$}
\author[4]{Jeffrey A. Kimbrel}
\author[4]{Megan M. Morris}
\author[4]{Xavier Mayali}
\author[1,5]{Cullen R. Buie$^\dagger$}
\affil[1]{Department of Mechanical Engineering, Massachusetts Institute of Technology}
\affil[2]{Institute for Data, Systems, and Society, Massachusetts Institute of Technology}
\affil[3]{Department of Statistics, University of California, Davis}
\affil[4]{Physical and Life Sciences Directorate, Lawrence Livermore National Laboratory}
\affil[5]{Department of Biological Engineering, Massachusetts Institute of Technology}
\date{}
\begin{document}

\maketitle
\def\thefootnote{*}\footnotetext{These authors also contributed equally to this work.}
\def\thefootnote{$\dagger$}\footnotetext{Correspondence: {\tt crb@mit.edu}}

\begin{abstract}
Multidimensional scaling (MDS) is a dimensionality reduction technique for microbial ecology data analysis that represents the multivariate structure while preserving pairwise distances between samples. While its improvement has enhanced the ability to reveal data patterns by sample groups, these MDS-based methods require prior assumptions for inference, limiting their application in general microbiome analysis.
In this study, we introduce a new MDS-based ordination, ``$F$-informed MDS,'' which configures the data distribution based on the $F$-statistic, the ratio of dispersion between groups sharing common and different characteristics. Using simulated compositional datasets, we demonstrate that the proposed method is robust to hyperparameter selection while maintaining statistical significance throughout the ordination process.
Various quality metrics for evaluating dimensionality reduction confirm that $F$-informed MDS is comparable to state-of-the-art methods in preserving both local and global data structures. Its application to a diatom-associated bacterial community suggests the role of this new method in interpreting the community’s response to the host.
Our approach offers a well-founded refinement of MDS that aligns with statistical test results, which can be beneficial for broader compositional data analyses in microbiology and ecology. This new visualization tool can be incorporated into standard microbiome data analyses.
\end{abstract}

\section*{Introduction}
Understanding microbial diversity has been advanced by the development of gene sequencing technology and multivariate data analytics, which together attempt to interpret the composition of microbial communities by targeting a conserved gene region in this phylogeny (16S rRNA gene) or by profiling all genes present (i.e., whole metagenome sequencing). 
Often referred to as the microbiome, the ecological structures obtained from these sequencing tools are distinguished from other biological data types in that they are compositional \cite{Greenacre21, Gloor17}, sparse \cite{Martino19}, and that their features are linked under a phylogenetic tree, providing additional genomic context \cite{Martin02, Armstrong22}.
Exploratory analysis of microbiome data \cite{peterson24, Armstrong22}, after preprocessing the sequencing reads into compositional data \cite{Lin20, Weiss17, McMurdie14, paulson13}, typically begins with visualizing the multivariate structure to identify data patterns, distinguish variations or remove outliers.
Heatmaps or bar plots can directly represent the relative abundances of each taxon or feature. However, for large datasets including high-throughput sequencing data, statistical ordination analysis is often required to detect patterns.

\begin{figure}[h!]
    \centering
    \includegraphics[width=5.2in]{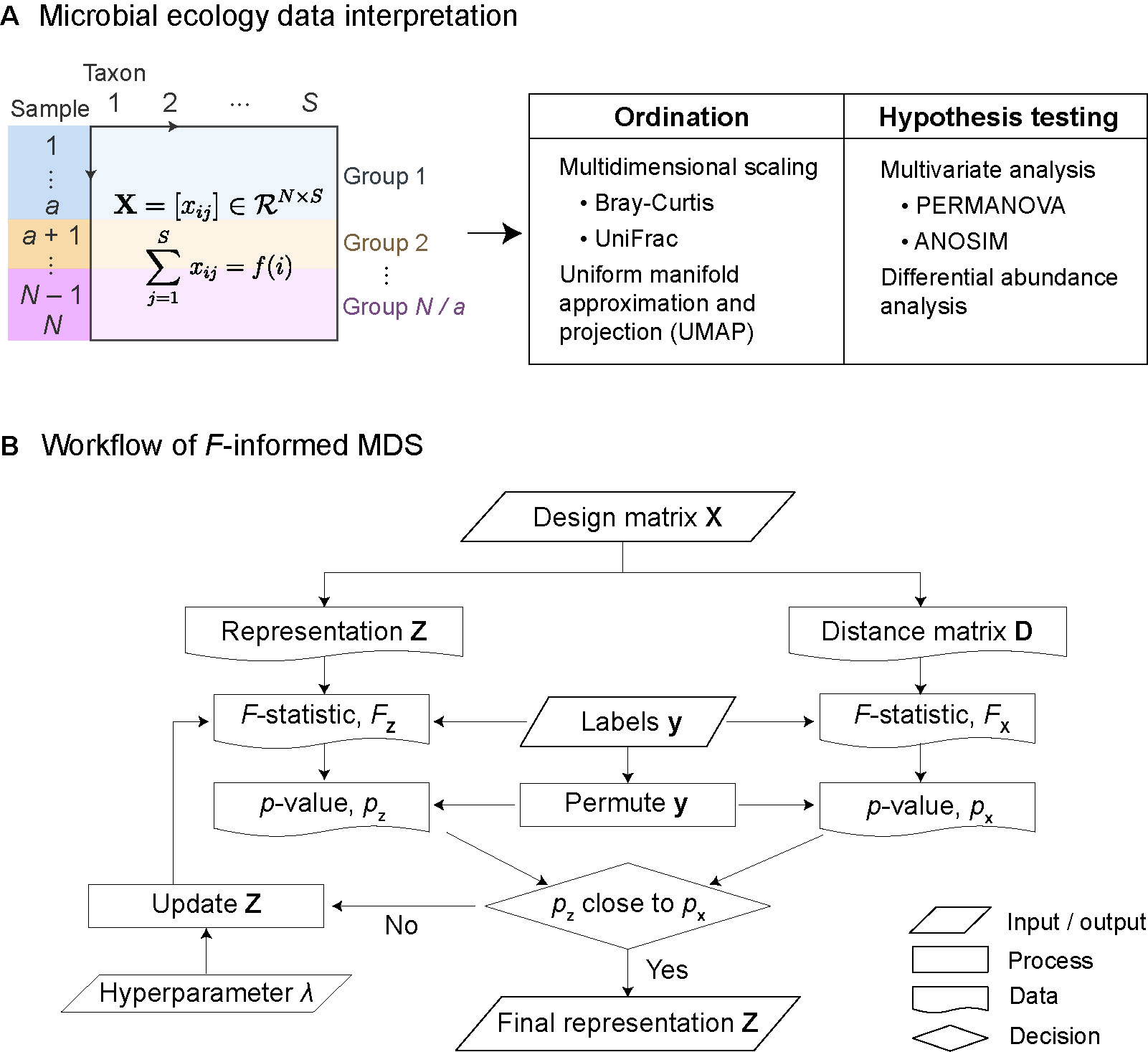}
    \caption{\textbf{\textit{F}-informed multidimensional scaling (MDS) for microbial ecology data analysis.} (a) Schematic overview of analyzing a dataset, represented as a design matrix $\mathbf{X}$ of $N$ samples with $S$ features or taxa. In a balanced design, samples are grouped by $a$ replicates of the same experimental condition. Note that compositionality requires that the summation of elements $x_{ij}$ across features is fixed or independent of a feature $j$ (e.g., 1). Exploratory analysis of $\mathbf{X}$ is performed through its ordination, followed by statistical inference, including hypothesis testing (e.g., PERMANOVA) or differential abundance analysis. (b) Computational process diagram of reducing the dimensionality of $\mathbf{X}$ using label set $\mathbf{y}$ for visualization, ensuring compliance with inference results based on the $F$-statistic.}
    \label{fig:intro}
\end{figure}

The most commonly used ordination method in ecology \cite{Jeganathan21} is multidimensional scaling (MDS), an ordination technique that represents data in a lower dimensional space (e.g., two-dimensional or 2D) while preserving its original distance structure (Figure \ref{fig:intro}A).
This is achieved by minimizing a stress function \cite{borg97a}, defined as the summation of all pairwise distances or dissimilarities. 
Using such a dissimilarity metric is essential for finding an appropriate 2D representation of these sparse or zero-inflated compositions \cite{gijbels13}, and different patterns have been observed depending on the choice of metric \cite{Armstrong22, Shi22}. Examples in microbiome analysis include Bray-Curtis \cite{Bray57} and Unifrac \cite{lozupone05, lozupone07}, which incorporate 16S rRNA gene-based taxonomic information as well as the compositional structure. Herein, we refer to these distance-based ordination methods to as metric MDS.
Applications of distance metrics have allowed MDS-based ordination to detect patterns and environmental gradients \cite{Kuczynski10}, or further address biologically confounding factors \cite{Shi20, Wang22}.
More recently, an alternative visualization tool for high-dimensional biological data, UMAP \cite{McInnes18}, has also been applied to microbiomes to cluster features associated with their sampling site within a host \cite{Armstrong21}.

Evaluating ordination or dimension reduction methods has been facilitated by establishing quantitative measures to assess their performance or appropriateness for specific objectives. 
These evaluations quantify the change or loss of information from the original structure using quality metrics widely employed in information visualization \cite{Bertini11,Gonzalez12,Espadoto21}, with applications in biological fields such as single-cell genomics \cite{Lause24} and transcriptomics \cite{Cui24, Narayan21}.
These metrics assess how well local patterns are preserved (e.g., trustworthiness, continuity \cite{Venna01}) or how few pairwise distances are distorted from the original structure (e.g., Shepard diagram, normalized stress \cite{borg05c}).
For microbiome data, however, the evaluations have primarily focused on the ability to identify group clustering on a case-by-case basis, using measures such as Rand index \cite{Shi22}, k-means clustering \cite{Kuczynski10, Wang22}, and the $F$-statistic \cite{Hawinkel19, Wang22}. Additionally, environmental gradient patterns have been analyzed using kernel- \cite{Randolph18} or rank-based \cite{Kuczynski10} regressions.

Notably, most ordination methods for ecological data analysis do not utilize data labels and instead rely on unlabeled or unweighted distances in their computations. Concurrently, an additional and independent differential abundance analysis is required to test for structural change and to correlate with treatment \cite{peterson24, Armstrong22}.
While metric MDS can still be used to detect the changes \cite{Borg05d}, its low-dimensional representation does not always capture every underlying pattern, even when statistical validation confirms the pattern in the original space. To address such cases, scree plot analysis is performed to determine the dimensionality at which the pattern information was preserved. 
However, if more than three dimensions are required, effectively visualizing the multivariate data becomes impractical.

This limitation has driven the development of enhanced methods that incorporate additional contextual information to improve pattern detection. One approach, broadly referred to as confirmatory MDS \cite{borg97b, ding18}, differentiates between sample groups by incorporating labels \cite{cox93, yang18}. This method is enabled by hyperparameter adjustments that optimize objective function \cite{yang18, Wang22, witten11}.
%where label-containing or -free terms are balanced \cite{yang18, Wang22, witten11}.
Adapted representations are generally accepted as long as they do not deviate significantly from the metric MDS representation \cite{borg97a}, particularly given that MDS optimizes a non-convex function and can yield different solutions \cite{Demaine21, zheng19}. 
For biological data applications, newer visualization methods have been introduced to address confounding factors by making certain assumptions about the relationship between errors and covariates \cite{Shi20} or by incorporating group labels into the latent structure \cite{Wang22, Lin16}. 
However, applying these methods to general microbiome datasets is limited by the assumptions regarding structural differences or data models, which are not always valid in real-world scenarios.

In this work, we propose a novel framework, $F$-informed MDS ($F$-MDS), designed to visualize statistical differences within an exploratory MDS setting by leveraging the $F$-statistic. 
Our approach aims to introduce minimal perturbations to the original MDS representation, preserving the global distance structure as much as possible. 
This generalizes previous label-based and model-based MDS methods by eliminating the need for prior assumptions about group structures or their differences. 
In the following sections, we demonstrate how $F$-informed MDS produces representations that are less dependent on hyperparameter choices and evaluate its performance in comparison to other dimension reduction tools.

\section*{Methods} \label{sec:methods}
In this section, we review the terminology and definitions of multidimensional scaling (MDS), along with an example of its supervised variant that incorporates sample group information.
Next, we propose a new MDS method and discuss multivariate analysis of variance via permutation (i.e., PERMANOVA \cite{anderson01}).
Finally, we introduce and justify quality metrics for evaluating dimension reduction tools and compare them with the proposed MDS approach.

\subsection*{Review of multidimensional scaling methods}
We consider a balanced design where an $i$-th sample, denoted as $\mathbf{x}_i \in \bbr^{S}$, is $S$-dimensional and associated to a discrete label $y_i \in \{0,1, \cdots r \}$ for every index $i=1,\cdots N$ (the number of samples, Figure \ref{fig:intro}A). 
A scalar, pairwise distance can be defined between $\mathbf{x}_i$ and $\mathbf{x}_j$, denoted as $d_{ij}\in \bbr_{+}^{N}$, by using a dissimilarity metric, e.g., Euclidean, Bray-Curtis \cite{Bray57} or UniFrac \cite{lozupone05}. 
Let us also denote $\mathbf{D} = [d_{ij}]$ the matrix or a set of the pairwise distances.

In the metric multidimensional scaling, a lower-dimensional representation $\mathbf{Z} = (\mathbf{z}_1, \cdots \mathbf{z}_N)\in \bbr^{2 \times N}$ is sought that best preserves the original, distance structure $\mathbf{D}$. 
This is enabled by minimizing an objective function $O_{\rm MDS}(\mathbf{Z})$ (termed the raw stress \cite{borg05c}),

\begin{equation}
    O_{\rm MDS}(\mathbf{Z})  = \frac{1}{2}\sum_{i,j} (d_{ij} - \| \mathbf{z}_i - \mathbf{z}_j \|_2 )^2.
\end{equation}

Known algorithms to minimize the raw stress include matrix decomposition via Nyström approximation \cite{Platt05}, divide-and-conquer \cite{Yang06, Qu15} for Euclidean distances, or more generally for non-convex function, greedy algorithms \cite{Demaine21} such as majorization \cite{deLeeuw88, borg97a} or stochastic gradient descent \cite{Kruskal64, zheng19}.
% a difference in the pairwise distance between the dimensions and summing the difference for all pairs, so-called 

While the metric MDS is unsupervised learning and does not require the label set $\mathbf{y}=(y_1,\cdots ,y_N)$, a supervised version of multidimensional scaling, e.g., SuperMDS \cite{witten11}, imposes an additional constraint on the representation by class groups.
The purpose is to classify samples by the labels as well as to compute the multidimensional scaling.
This is enabled by adding a confirmatory/constraining term to the raw stress, 
\begin{equation}
\begin{split}
&O_{\rm SMDS}(\mathbf{Z}) = \\
&\quad (1-\alpha)\cdot \underbrace{\frac{1}{2}\sum_{i,j} (d_{ij} - \| \mathbf{z}_i - \mathbf{z}_j \|_2 )^2}_\text{raw stress} + 
\alpha \underbrace{\sum_{i,j:y_{j}>y_{i}} (y_j - y_i) \sum_{k=1}^2 \left(\frac{d_{ij}}{\sqrt{2}}-(z_{jk} - z_{ik})\right)^2}_\text{confirmatory term},
\end{split}
\end{equation}
where the confirmatory term involves the groups set, balanced by the hyperparameter $\alpha$ that controls the degree of classification.  
Minimizing $O_{\rm MDS}(\mathbf{Z})$ can locate the representation points closer to each other when they are within the same group.
Selection of the hyperparameter is carefully guided by the data structure, so that the process avoids unnecessary group distinctions during the visualization.

\subsection*{Proposal of multidimensional scaling informed by pseudo $F$-statistic}
% \textcolor{red}{[Describe the outline of this section here.]}
In this section we introduce a multivariate test statistic based on samples dispersion (i.e., \textit{F}-statistic) and propose a revised MDS that utilizes the test statistic.

\paragraph*{Permutational multivariate analysis of variance.}
Testing for group differences in multivariate data is commonly performed by calculating the $F$-statistic, which compares inter- and intra-group variances based on pairwise distances.
However, a standard $F$-test assumes that observations are normally distributed with equal variance, an assumption that does not hold for most compositional or zero-inflated data.
Instead, a ``pseudo'' $F$-statistic has been defined \cite{clarke93}, by denoting $\epsilon_{ij}$ the $\1\{y_i=y_j\}$ (indicator function) to write
\begin{equation} \label{eq:pseudof}
F = \frac{\sum_{i,j}d_{ij}^2 - 2\sum_{i,j}\epsilon_{ij}d_{ij}^2}{2\sum_{i,j}\epsilon_{ij}d_{ij}^2}\cdot (N-2).
\end{equation} 

Using the pseudo-$F$'s, an empirical distribution can be constructed by permuting the labels for a large sized dataset \cite{holmes96, gijbels13}.
Specifically, a new $F$-ratio denoted as $F^{\pi}$ is calculated from the newly permuted label set $[y_i]^{\pi}$ and the calculation is repeated over $K$ independent permutations, $\pi_1 , \cdots ,\pi_K$. This pseudo-$F$-distribution yields a $p$-value, the procedure known as the permutational multivariate analysis of variance (PERMANOVA) \cite{anderson01}:
\begin{equation}
p = \frac{1}{K}\,\sum_{k=1}^{K}\1\{F^{\pi_k}\geq F\}.
\end{equation}

\paragraph*{Proposal of $F$-informed MDS.}
To incorporate hypothesis testing into MDS visual analysis, we take a weakly supervised approach using the following steps (Figure \ref{fig:intro}B).
First, a two-dimensional representation is initialized by computing metric MDS, which is unsupervised and label-free.
Next, the representation points are iteratively adjusted to ensure that the $p$-values in 2D space closely match those in the original structure.
This is enabled by optimizing an objective function with a new confirmatory term added to the raw stress,
\begin{equation} \label{eq:fmds}
    O_{\rm FMDS}(\mathbf{Z}) \sim
    \underbrace{\frac{1}{2}\sum_{i,j} (d_{ij} - \| \mathbf{z}_i - \mathbf{z}_j \|_2 )^2}_\text{raw stress} +
    \lambda\cdot \underbrace{\left|F_\mathbf{z} - f_\mathbf{z}(F_\mathbf{x})\right|}_\text{confirmatory term},
\end{equation}
in which the confirmatory term can minimize a difference between $p$-values from the $F$-distributions represented at each dimensionality, i.e., original and the two-dimension (herein denote them as $F_\mathbf{x}$ and $F_\mathbf{z}$, respectively). 
In detail, after noticing that the distributions can vary depending on the dimension, we compensated for the difference by introducing a local regression function $f_\mathbf{z}$, that maps between the two $F$-ratios, each `pre-ordered' by size based on permuted labels.
In other words, a local regression is performed using two ordered $F$-ratios, $(F_\mathbf{x}, F_\mathbf{z})$, where each element in the tuple is computed from two independent, permuted sets, $\mathbf{y}^{\pi_1}$ and $\mathbf{y}^{\pi_2}$ (Algorithm \ref{alg:mapping}). Details on the mapping function and the derivation of an exact $O_{\rm FMDS}$ are explained in \ref{sec:app1}.

\begin{algorithm}[h]
\caption{Mapping function $f_\mathbf{z}$.}
\label{alg:mapping}
\begin{algorithmic}[1]
\Require{
\begin{tabular}{@{}ll}
    $\mathbf{Z} \in \bbr^{N \times 2}$ & : representation \\
    $\mathbf{D} \in \bbr_{+}^{N \times N}$ & : pairwise distance \\
    $\mathbf{y} = \{0, 1,\cdots r\}^N$ & : labels or groups
\end{tabular}
}
\Ensure{
\begin{tabular}{@{}ll}
    $f_\mathbf{z}:\bbr\rightarrow\bbr$ & : mapping function    
\end{tabular}
}
\State$\mathcal{L}_\mathbf{x} \Leftarrow$ empty list
\State$\mathcal{L}_\mathbf{z} \Leftarrow$ empty list
\For{$1\leq i \leq 999$}
    \State$\mathbf{y}^{\pi_1} \Leftarrow$ random permutation on $\mathbf{y}$
    \State$\mathbf{y}^{\pi_2} \Leftarrow$ random permutation on $\mathbf{y}$
    \State$F^{\pi_1}_\mathbf{x} \Leftarrow$ pseudo-$F$ using $\mathbf{D}$ and $\mathbf{y}^{\pi_1}$
    \State$F^{\pi_2}_\mathbf{z} \Leftarrow$ pseudo-$F$ using $\mathbf{Z}$ and $\mathbf{y}^{\pi_2}$
    \State$\mathcal{L}_\mathbf{x} \Leftarrow$ append $F^{\pi_1}_\mathbf{x}$ to $\mathcal{L}_\mathbf{x}$
    \State$\mathcal{L}_\mathbf{z} \Leftarrow$ append $F^{\pi_2}_\mathbf{z}$ to $\mathcal{L}_\mathbf{z}$
\EndFor
\State$\mathcal{L}_\mathbf{x} \Leftarrow$ sort $\mathcal{L}_\mathbf{x}$ by an increasing order
\State$\mathcal{L}_\mathbf{z} \Leftarrow$ sort $\mathcal{L}_\mathbf{z}$ by an increasing order
\State$f_\mathbf{z} \Leftarrow$ local regression from $\mathcal{L}_\mathbf{x}$ to $\mathcal{L}_\mathbf{z}$
\end{algorithmic}
\end{algorithm}

\paragraph*{Majorization algorithm.}
In our problem setting, we implemented the majorization (or Majorization-Minimization) algorithm to minimize the $F$-MDS objective (Equation \ref{eq:fmds}). First, an exact form of $O_{\rm FMDS}(\mathbf{Z})$ was derived to match the physical dimension of each term (\ref{sec:app1}),
\begin{equation}
    O_{\rm FMDS}(\mathbf{Z}) = \sum_{i,j} (d_{ij} - \| \mathbf{z}_i - \mathbf{z}_j \|_2 )^2 + \lambda\left|\sum_{i,j} \left[1- 2\epsilon_{ij} \left(1+\frac{f_\mathbf{z}(F_\mathbf{x})}{N-2}\right)\right] \|\mathbf{z}_i - \mathbf{z}_j\|_2^2\right|.
\end{equation}
Next, as similarly described in \cite{borg05c, witten11}, the majorization is applied to seek an optimal $k$-th point for every $k=1,\cdots N$, while keeping other points except for $\mathbf{z}_{k}$ fixed,
\begin{equation}
    \mathbf{z}_k^* = \argmin_{\mathbf{z}_k}O_{\rm FMDS}(\mathbf{Z} | \mathbf{z}_1,\cdots \mathbf{z}_{k-1}, \mathbf{z}_{k+1}, \mathbf{z}_N).
    \label{eq:seek_z}
\end{equation}
It can be analytically shown that Equation \ref{eq:seek_z} expression can be majorized to a quadratic expression in terms of $\mathbf{z}_k$, where the quadratic is further minimized by taking a derivative with respect to each element of $\mathbf{z}_k$ (denoted as ${z}_{ks},\ s=1,\cdots ,r$). An update rule was established by solving for ${z}_{ks}$ after setting the derivative to zero. Detailed derivation of the expression and the update rule for $\mathbf{Z}$ are described in \ref{sec:app2}, and the whole procedure is summarized in Algorithm \ref{alg:mm}.

\begin{algorithm}[h]
\caption{Majorization for computing $F$-MDS.}
\label{alg:mm}
\begin{algorithmic}[1]
\Require{
\begin{tabular}{@{}ll}
    $\lambda \in \bbr_{+}$ & : hyperparameter \\
    $\mathbf{D} \in \bbr_{+}^{N \times N}$ & : pairwise distance \\
    $\mathbf{y} = \{0, 1,\cdots r\}^N$ & : labels or groups
\end{tabular}
}
\Ensure{
\begin{tabular}{@{}ll}
    $\mathbf{Z} \in \bbr^{N \times 2}$ & : 2D configuration
\end{tabular}
}
\State$F_\mathbf{x} \Leftarrow$ pseudo-$F$ using $\mathbf{D}, \mathbf{y}$ (Equation \ref{eq:pseudof})
\For{$1\leq t \leq T$}
\For{$1\leq k \leq N$}
    \State$\delta(\mathbf{Z}) \Leftarrow$ sign of confirmatory term (Equation S13)
    \State$f_\mathbf{z}(F_\mathbf{x}) \Leftarrow$ mapping function (Algorithm \ref{alg:mapping})
    \State$\mathbf{z}_k \Leftarrow$ update $\mathbf{z}_k$ with $\delta(\mathbf{z})$ and $f_\mathbf{z}(F_\mathbf{x})$  (Equation S18)
\EndFor
\EndFor
\end{algorithmic}
\end{algorithm}

\subsection*{Dataset and evaluation}
\paragraph*{Description of simulation study and real dataset.}
Simulation studies were performed using randomly generated four-dimensional datasets to illustrate cases where metric MDS fails to reveal group-based patterns in a two-dimensional representation. The datasets were designed to follow multivariate normal distributions, with each group having different means but the same covariance matrix (Equation \ref{eq:binary_distrib}-\ref{eq:ternary_distrib}). The objective of these simulations was to demonstrate the effectiveness of the proposed method in cases where group differences are not distinguishable.
However, if the difference is significant and visible, then metric multidimensional scaling alone is sufficient and does not require the majorization process outlined earlier. 

Specifically for a binary dataset, triplicates of 100 samples were randomly drawn from truncated normal distributions \cite{Adams22} (denoted as $\ctN (\cdot, \cdot)$) with following parameters,
\begin{equation}
\begin{aligned}
    \mathbf{w}_i &\sim 
    \begin{cases}
    \ctN \left(\bm\upmu + \bm\updelta, \Sigma_{\mathbf{w}}\right), & 1\leq i\leq 50 \\
    \ctN \left(\bm\upmu - \bm\updelta, \Sigma_{\mathbf{w}}\right), & 51\leq i\leq 100,
    \end{cases} \\    
    \bm\upmu = \frac{1}{4}
    \begin{bmatrix}
    1 \\ 1 \\ 1 \\ 1
    \end{bmatrix}, \ 
    &\bm\updelta = \frac{1}{20\sqrt{2}}
    \begin{bmatrix}
    1 \\ -1 \\ 0 \\ 0
    \end{bmatrix}, \ 
    \Sigma_{\mathbf{w}} = \frac{1}{100}
    \begin{bmatrix}
    0.01 & 0& 0& 0 \\ 0& 4& 0& 0 \\ 0& 0& 4& 0 \\ 0& 0& 0& 1
    \end{bmatrix}.
    \label{eq:binary_distrib}
\end{aligned}
\end{equation}
A compositionality characteristic was additionally represented by applying a total sum scaling to the multivariate data, i.e., dividing a sample components with their sum,
\begin{equation}
    \mathbf{x}_i = \frac{\mathbf{w}_i}{\sum_{j=1}^N\, w_{ji}}.
    \label{eq:TSS}
\end{equation}
And for ternary dataset, 150 samples were simulated by following,
\begin{equation}
\begin{aligned}
    \mathbf{x}_i &\sim 
    \begin{cases}
    \cN \left(\bm\upmu_1, \Sigma_{\mathbf{x}}\right), & 1\leq i\leq 50 \\
    \cN \left(\bm\upmu_2, \Sigma_{\mathbf{x}}\right), & 51\leq i\leq 100 \\
    \cN \left(\bm\upmu_3, \Sigma_{\mathbf{x}}\right), & 101\leq i\leq 150,
    \end{cases} \\    
    \bm\upmu_1 = 
    \begin{bmatrix}
    0 \\ 0 \\ 0 \\ 0
    \end{bmatrix}, \ 
    \bm\upmu_2 &=
    \begin{bmatrix}
    0 \\ 0 \\ 2 \\ 0
    \end{bmatrix}, \ 
    \bm\upmu_3 =
    \begin{bmatrix}
    0 \\ 0 \\ 1 \\ \sqrt{3}
    \end{bmatrix}, \ 
    \Sigma_{\mathbf{x}} = 
    \begin{bmatrix}
    5& 0& 0& 0 \\ 
    0& 5& 0& 0 \\ 
    0& 0& 1& 0 \\ 
    0& 0& 0& 1
    \end{bmatrix}.
    \label{eq:ternary_distrib}
\end{aligned}
\end{equation}

We also evaluated $F$-informed MDS using bacterial community data associated with or sampled from hosts, including a photosynthetic diatom \cite{kim22} and the human gut, the latter used for studying liver cirrhosis \cite{Qin14} or type 2 diabetes (T2D) \cite{Karlsson13, Qin12}. 
For the algal-associated microbiome, the dataset consisted 36 balanced microbial community samples cultured with or without the microbial host (alga \textit{Phaeodactylum tricornutum}). The community comprised 72 bacterial taxa as identified by amplicon sequence variants (ASV) of the 16S rRNA gene, PCR-amplified. ASV counts were obtained through Illumina MiSeq system and subsequently converted into relative abundances using cumulative sum scaling (CSS) \cite{paulson13}. A phylogeny-based metric (weighted Unifrac \cite{lozupone07}) and compositional data were used to calculate pairwise distance between samples, forming the distance matrix $\mathbf{D}$.
For the human gut microbiome, two datasets generated from shotgun metagenomic sequencing were retrieved from a publicly available repository \cite{Pasolli16}, where reads were merged at the genus level \cite{reiman20}. Detailed reproduction and processing steps for these datasets are described in \ref{sec:app3}.

\paragraph*{Performance evaluation metrics.}
To quantify the performance of dimension reductions, four known \cite{Espadoto21} and two new quality metrics were defined and jointly used. Trustworthiness and continuity, among the first four metrics, measure how closely neighborhood points are preserved from the original ($S$-) to two-dimensional representation, or vice versa, defined in \cite{Venna01} and expressed as
\begin{equation}
\begin{aligned}    
    \text{Trustworthiness} &= 1 - \frac{2}{Nk(2N-3k-1)}\sum_{i=1}^N\sum_{j\in U_k (i)}\, (r_{ij}-k) \\
    \text{Continuity} &= 1 - \frac{2}{Nk(2N-3k-1)}\sum_{i=1}^N\sum_{j\in V_k (i)}\, (\hat{r}_{ij}-k),
    \label{eq:def_venna}
\end{aligned}
\end{equation}
where $k$ is the size of neighborhood of interest, $r_{ij}$ and $\hat{r}_{ij}$ respectively are the distance-based ranks of $j$-th point from $i$-th point under the original and/or two-dimensional space. 
Sets $U_k (i)$, $V_k (i)$ are constructed by following, $U_k = \hat{C_k}\cap C_k^c$ and $V_k = {C_k}\cap \hat{C_k}^c$, where $C_k(i)$ and $\hat{C_k}(i)$ are the set of $k$ points closest to $i$-th point in the original and two-dimensional space, respectively. Two different $k$ values, corresponding to $\sim 8\%$ and $75\%$ of data size, were applied to evaluate how local and global information is preserved in simulated and real datasets.
Trustworthiness and Continuity were computed using an R package \texttt{dreval} (v.0.1.5 \cite{Soneson24}).

Additionally, we quantified the degree of deviation of the representation $\mathbf{Z}$ from the original distance $\mathbf{D}$ by calculating two metrics; the first was the ``normalized'' stress or Stress-1 \cite{borg05c},
\begin{equation}
\text{Stress-1} = \frac{\sum_{i,j} (d_{ij} - \| \mathbf{z}_i - \mathbf{z}_j \|_2)^2}{\sum_{i,j} \| \mathbf{z}_i - \mathbf{z}_j \|_2^2},
\label{eq:mds_stress1}
\end{equation}
and the second using a Shepard diagram \cite{dexter18} and its Pearson correlation coefficient. Both metrics provided how much distortion from the original global structure had occurred in the representation from each reduction method.

Two new metrics were also defined and used, which we termed as $F$-correlation and $F$-rank-ratio,
\begin{equation}
\begin{aligned}
    F\text{-correlation} &= \text{corr}(F_\mathbf{x}^\pi, F_\mathbf{z}^\pi) \\
    F\text{-rank-ratio} &= \frac{1-p_\mathbf{z}}{1-p_\mathbf{x}} = \frac{\sum_{k=1}^K\,\1\{F_\mathbf{z}^{\pi_k} < F_\mathbf{z}\}}{\sum_{k=1}^K\,\1\{F_\mathbf{x}^{\pi_k} < F_\mathbf{x}\}},
    \label{eq:def_f_metrics}
\end{aligned}
\end{equation}
with $\text{corr}(\cdot,\cdot)$ denoting the Pearson coefficient between two random variables and superscript $\pi$ indicating a permutation of group labels $\mathbf{y}$.

Experiments were performed using R via RStudio either on Apple M3 chip with 16 GB of RAM, or on 128-thread IBM POWER9 CPU with 256 GB of RAM at the MIT Office of Research Computing and Data.

% Results and Discussion can be combined; please write results in past tense.
\section*{Results}
\subsection*{$F$-informed MDS is robust to a choice of hyperparameter.}
We first characterized the behavior of $F$-informed MDS by observing changes in the MDS representation across a range of hyperparameters. The changes were recorded after each update of the representation $\mathbf{Z}$ by computing its objective function, $O_{\rm FMDS}(\mathbf{Z})$, and the $p$-value from PERMANOVA testing of $\mathbf{Z}$ ($p_\mathbf{z}$, Figure \ref{fig:intro}). The $p$-value was chosen as a tracking measure instead of the pseudo $F$-statistic because its distribution varied depending on the hyperparameter $\lambda$ (\ref{sec:app1}). A binary, balanced dataset of 100 random samples was simulated and normalized using total sum scaling, where each sample group followed a four-dimensional truncated normal distribution with the same covariance but different means (\ref{sec:methods}). A scatter matrix of the data, re-based using orthonormal vectors, confirmed that the samples were compositional. Each group exhibited an elliptical 3D distribution with a distinct mean positioned along the shorter axis, but the group difference was indistinguishable due to high variance along the two longer axes (\ref{fig:sim_data_2d}). Indeed, a two-dimensional representation using metric MDS with Euclidean distance did not indicate a statistically significant group difference ($p_\mathbf{z}=0.645$), whereas the original 4D data showed a significant difference ($p_\mathbf{x}=0.004$).

\begin{figure}[h!]
    \centering
    \includegraphics[width=5.2in]{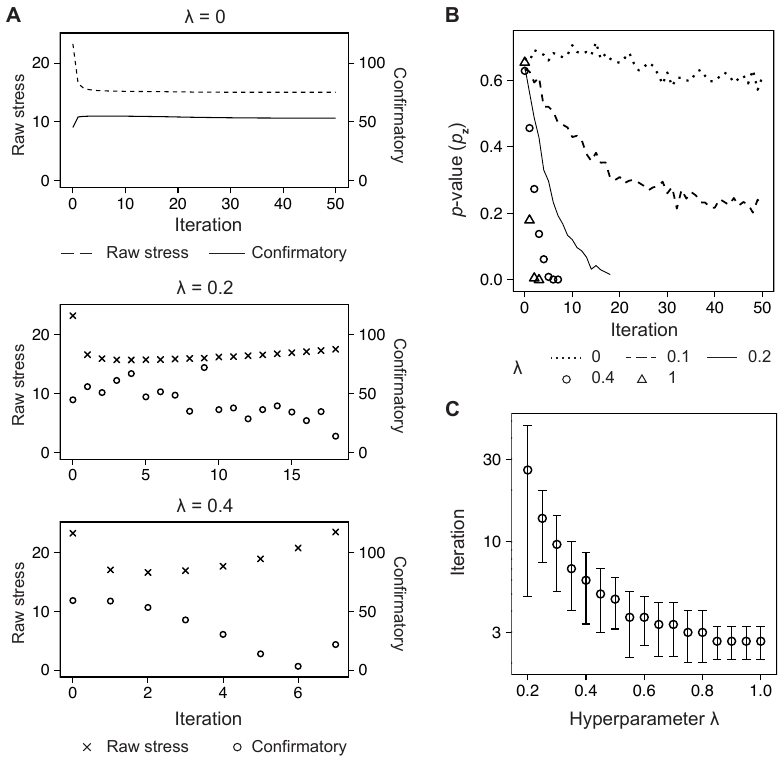}
    \caption{\textbf{Majorization algorithm optimizes the \textit{F}-informed MDS objective function.} (A) Changes in each term comprising the objective function (Equation \ref{eq:fmds}), including raw stress and confirmatory terms, are plotted against the number of iterations for different hyperparameter values $\lambda$. (B) The PERMANOVA $p$-value under two-dimensional representation ($p_\mathbf{z}$) is plotted against the iteration number and $\lambda$. (C) The log-scaled iteration number is plotted against the hyperparameter, ranging between 0.2 and 1. Three compositional datasets were simulated, each with two groups following truncated normal distributions with the same covariance but different means (\ref{fig:sim_data_2d}). Error bars represent the standard deviation of triplicates drawn from the distributions.}
    \label{fig:sim_iter}
\end{figure}

Tracking changes in the $F$-MDS representation over iterations revealed that a minimal hyperparameter value was required for the majorization algorithm to minimize the objective function. In one simulated dataset, we observed a gradual decrease in $p_\mathbf{z}$-value and $O_{\rm FMDS}(\mathbf{Z})$ with each iteration, as expected. The algorithm converged in fewer than 20 iterations when $\lambda\geq0.2$ (Figure \ref{fig:sim_iter}A). In our setting, termination occurred when $p_\mathbf{z}$ was sufficiently close to $p_\mathbf{x}$ (e.g., difference less than 0.01, Figure \ref{fig:intro}B). This trend was further examined by repeating the experiment across triplicates of the simulated data, confirming that $\lambda=0.2$ was the minimal value required for convergence and optimization (Figure \ref{fig:sim_iter}B). To ensure that the majorization form of $O_{\rm FMDS}(\mathbf{Z})$ remained convex with a positive quadratic coefficient, the hyperparameter value was set below unity, ranging from [0,1] (\ref{sec:app2}).

\begin{figure}[h!]
    \centering
    \includegraphics[width=5.2in]{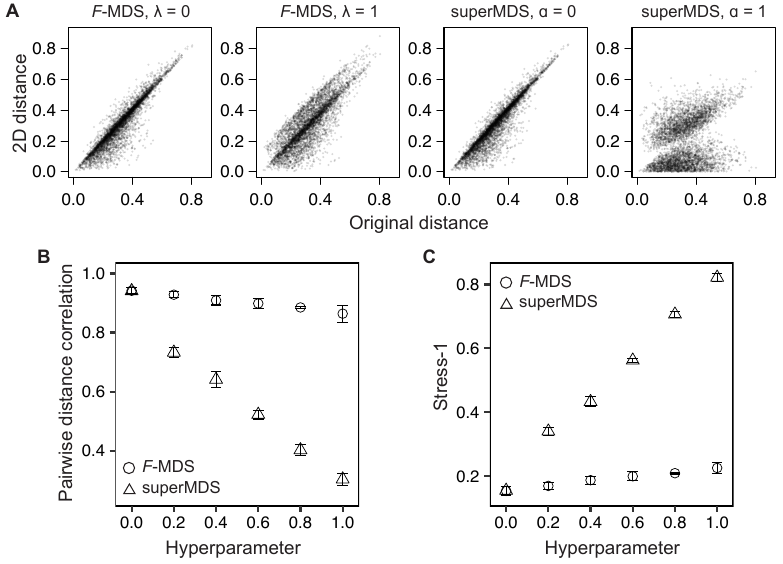}
    \caption{\textbf{\textit{F}-informed MDS is robust to the choice of hyperparameters.} (A) Shepard plots of pairwise distances using $F$-informed MDS ($F$-MDS) and supervised MDS (superMDS) with both zero (i.e., metric MDS) and nonzero hyperparameter settings ($\lambda=1$). For each dimensionality, distances were measured using weighted UniFrac (original) and Euclidean (2D). The pairwise distances from the original and two-dimensional representations were then used to calculate (B) their Pearson correlation coefficient and (C) normalized stress (Stress-1). Error bars represent the standard deviation of calculations from triplicate simulation datasets.}
    \label{fig:mds_eval_sim}
\end{figure}

After confirming the convergence of the objective with different hyperparameters, we next asked whether these values for $F$-MDS impacted its 2D representation. The visualizations were assessed using Shepard diagram and Stress, two quality metrics commonly used to evaluate how well an MDS representation preserves the distance structure from the original dimension (see \ref{sec:methods}). Supervised MDS (superMDS \cite{witten11}) was chosen as a benchmark ordination method because its objective function is similarly controlled by a single hyperparameter $\alpha$ within the same range $[0,1]$, ensuring convergence of its majorization algorithm. Shepard diagram analysis of these MDS methods showed that pairwise Euclidean distances deviated from the original structure in the two-dimensional representation to a greater extent when a nonzero hyperparameter was used. Interestingly, however, the change was more subtle in $F$-MDS (Figure \ref{fig:mds_eval_sim}A, \ref{fig:shepard_sim_all}). 

To further compare structural deviations, we calculated the Pearson correlation coefficient from the Shepard plot as well as the normalized stress (i.e., Stress-1, Equation \ref{eq:mds_stress1}). The pairwise distances correlation consistently remained above 0.85 for all tested values of $\lambda$ in $F$-MDS, whereas superMDS showed a decrease in correlation coefficients to $\sim\!0.3$ as $\alpha$ increased (Figure \ref{fig:mds_eval_sim}B). Similarly, the increase in Stress-1 due to a higher hyperparameter was 9.3-fold greater in superMDS than in $F$-MDS (Figure \ref{fig:mds_eval_sim}C). Together with previous results, these findings confirm that the proposed method retains the original distance information as effectively as metric MDS while being less dependent on the hyperparameter. At the same time, it successfully rearranges the 2D structure so that the groups become statistically distinct. Additionally, since the iterations required to compute $F$-MDS are minimized and reach a plateau at $\lambda\approx1$ (Figure \ref{fig:sim_iter}C), we suggest that the majorization algorithm provides an effective strategy for hyperparameter selection, ensuring high-quality performance while reducing computational cost.

\subsection*{Using different quality metrics confirms the high performance of $F$-MDS.}
Our earlier results demonstrate that $F$-informed MDS effectively represented the distance structure in 2D space. Next, we sought to generalize our statistical approach by comparing $F$-MDS to other dimension reduction methods, as recent examples have shown these methods can be also used to visualize microbiome data. To evaluate each ordination more comprehensively, we introduced four quality metrics alongside those previously discussed (i.e., pairwise distance correlation and Stress-1). The first two, Trustworthiness and Continuity \cite{Venna01}, measure the degree to which local neighborhood structures are preserved in the 2D representation (Equation \ref{eq:def_venna}). While these metrics do not require group labels, they have been used in previous studies to evaluate gene expression analysis methods \cite{Nikkila02, Kaski03}. The other two metrics assess how closely pseudo $F$-distributions derived from the original multidimensional structure are reflected in the 2D representations (Equation \ref{eq:def_f_metrics}). First, $F$-correlation measures the Pearson correlation between two $F$-ratios: one computed from the original data's distance structure and the other from 2D representation using permuted labels, as described in \cite{Serviss17}. Each pair of $F$-ratios was obtained from single permutation, and $K=500$ permutations were performed to compute the $F$-correlation coefficient. A high $F$-correlation value indicates that the ordination method successfully reconstructs a similar dispersion pattern in the lower-dimensional space. Second, the $F$-rank-ratio was defined as the ratio of the reversed rank of $F$ from unpermuted labels to the rank of an ordered $F$-set obtained from $K$ permutations. This ratio was calculated for both the original and 2D representations. An $F$-rank-ratio close to 1 indicates that the ordination accurately preserves the statistical significance of group differences as presented in the original structure. These two metrics were used together to evaluate whether an ordination method retains the original pseudo $F$-distribution while also preserving statistical differences between groups.

\begin{figure}[h!]
    \centering
    \includegraphics[width=5.2in]{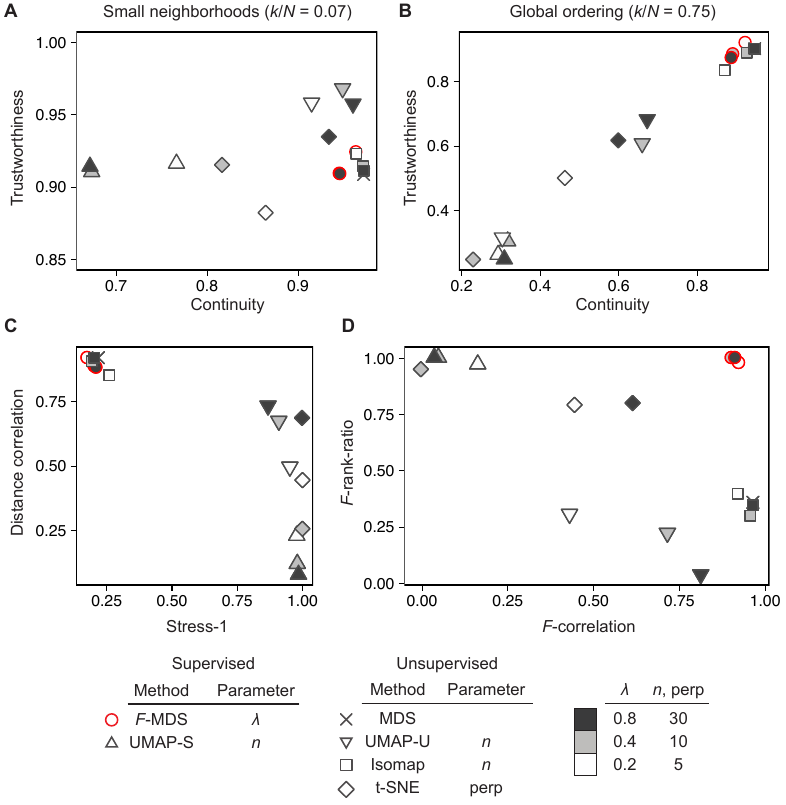}
    \caption{\textbf{Different quality metrics confirm consistent preservation of the simulated data pattern with \textit{F}-informed MDS.} Seven dimension reduction methods were evaluated to test their preservation of (A) local and (B) global structure by calculating trustworthiness and continuity using two nearest-neighbor numbers, $k=7$ (local), $k=75$ (global). The methods were also evaluated using (C) global distortion metrics (Stress-1 and Shepard diagram correlation) and (D) statistical inference metrics, including the ratio in statistical significance ($F$-rank-ratio) and correlation in $F$-ratios ($F$-correlation) using a randomly permuted label set. The following hyperparameters were applied to each method: $\lambda$ ($F$-MDS), $\alpha$ (SMDS), number of neighbors (UMAP, supervised (-S) and unsupervised (-U)), perplexity (t-SNE), and the number of shortest dissimilarities (Isomap).}
    \label{fig:eval_simul}
\end{figure}

\paragraph*{Simulated dataset.} \label{sec:eval_sim}
Using six quality metrics, the performance of $F$-MDS was evaluated and compared to other ordinations, including metric MDS, 
UMAP \cite{McInnes18}, t-SNE \cite{maaten08}, and Isomap \cite{tenenbaum00}. Simulated dataset was first used to compute the metrics across a range of hyperparameters. We observed that for local neighborhoods preservation, i.e., 7\% of data size, unsupervised UMAP (UMAP-U) achieved the highest trustworthiness score, followed by t-SNE, Isomap and ($F$-)MDS (Figure \ref{fig:eval_simul}A). While both unsupervised and supervised UMAP exhibited high trustworthiness ($>0.9$), the latter showed lower continuity, suggesting that additional constraints were introduced into the 2D representation when group labels were incorporated. In contrast, $F$-MDS achieved the highest continuity among all ordination methods while maintaining a trustworthiness score above 0.9. Overall, UMAP-U best preserved local neighborhood information for these simulated data, while our method demonstrated comparable performance with lower reliance on the hyperparameter. 

For global structure preservation, MDS-based ordinations (e.g., metric MDS, $F$-MDS, Isomap) were the top performers, as expected, when evaluated using the same two metrics but with broader neighborhoods (75\% data size, Figure \ref{fig:eval_simul}B) or distance-based metrics such as pairwise distance correlation (Shepard diagram) or Stress-1. The latter two metrics revealed that graph-based methods such as UMAP and t-SNE consistently resulted in higher Stress-1 values than MDS methods (Figure \ref{fig:eval_simul}C). This finding confirms that $F$-MDS does not significantly alter the global data structure compared to metric MDS. Finally, applying $F$-based metrics to evaluate these ordination methods showed their varying capabilities in representing distance structure in a lower dimension (Figure \ref{fig:eval_simul}D, \ref{fig:f_corr_sim_all}). Methods that incorporate group labels, such as $F$-, superMDS and supervised UMAP (UMAP-S), had $F$-rank-ratio values close to unity, indicating that group differences were accurately preserved in their 2D representations. In contrast, $F$-correlation was higher in unsupervised methods and those designed to retain distance structures, such as metric MDS, Isomap, and $F$-MDS, consistent with earlier observations from pairwise distance correlation.

\paragraph*{Real dataset using bacterial communities.}
We next analyzed a real microbial community as another example to verify the performance of these ordination methods. Each sample presented a compositional structure of relative abundances based on 16S rRNA gene expression, identifying 72 bacterial taxa collected from mesocosms of the alga \textit{P. tricornutum} \cite{Samo18, Kimbrel19}. Thirty-six community samples, half of which were co-cultured with the host \textit{P. tricornutum} and the other half without, were retrieved and re-analyzed from previous work \cite{kim22}. This dataset was chosen for our main analysis and evaluation because metric MDS did not detect a group difference ($p_\mathbf{z}\approx0.5$), whereas their original structures were statistically more different ($p_\mathbf{x}<0.1$). Other benchmark datasets, e.g., human gut microbiome, did not require additional iterations for $F$-MDS, as metric MDS already showed the significant differences ($p_\mathbf{z}= 0.004$, cirrhosis; 0.004, type 2 diabetes). For all datasets, weighted Unifrac \cite{lozupone07} was chosen as the distance metric to obtain the distance structure. Six quality metrics were again used to compare across seven dimension reduction methods, including one based on self-supervised learning (SimCLR \cite{chen20}), to further explore its capability of arranging the algal microbiome data for interpretation (\ref{sec:app4}). 

\begin{figure}[h!]
    \centering
    \includegraphics[width=5.2in]{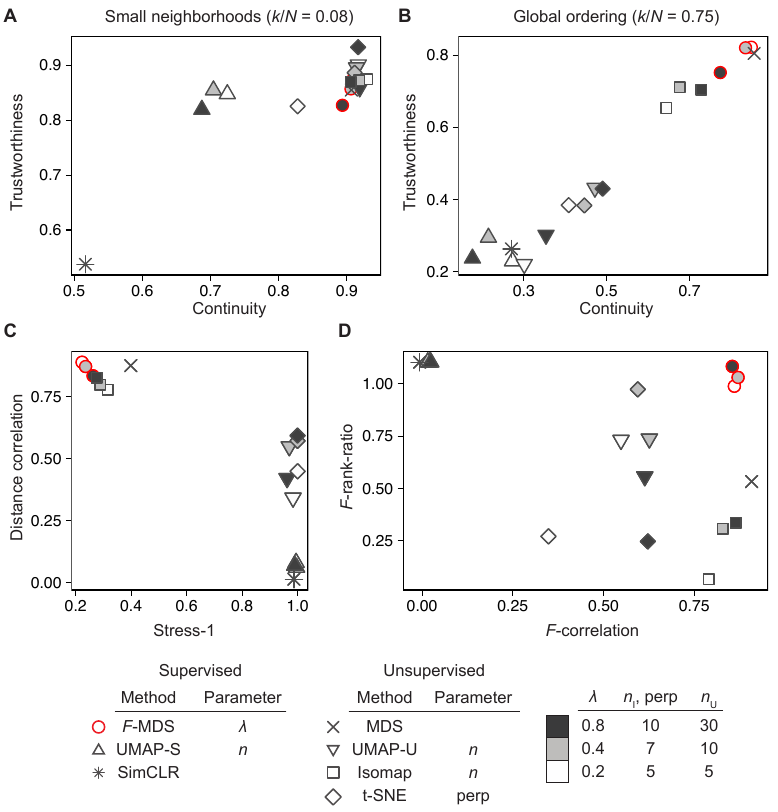}
    \caption{\textbf{Different quality metrics confirm consistent preservation of the algal microbiome pattern with \textit{F}-informed MDS.} Seven dimension reduction methods were evaluated for their preservation of (A) local and (B) global structure by calculating trustworthiness and continuity using two nearest neighbor numbers $k=3$ (local), $k=27$ (global). The methods were also assessed based on (C) global distortion metrics (Stress-1 and Shepard plot correlation) and (D) statistical inference metrics, including the ratio of $p$-values ($F$-rank-ratio) and correlation in $F$-ratios ($F$-correlation) using randomly permuted label set. The following hyperparameters were applied to each method: $\lambda$ ($F$-MDS), $\alpha$ (SMDS), number of neighbors (UMAP, supervised (-S) and unsupervised (-U)), perplexity (t-SNE), and the number of shortest dissimilarities (Isomap).}
    \label{fig:eval_real}
\end{figure}

Similarly, UMAP-U and Isomap achieved the highest scores for preserving local neighborhoods, slightly outperforming $F$-MDS as measured by trustworthiness. In contrast, other supervised or label-incorporating methods had continuity scores that were 22-44\% lower than Isomap (Figure \ref{fig:eval_real}A). For global structure preservation, however, $F$-MDS, metric MDS and Isomap were the top performers based on the two metrics as well as the pairwise distance correlation (Figure \ref{fig:eval_real}B,C, \ref{fig:shepard_alga_all}). Finally, statistical evaluations showed that $F$-MDS was again effective in retaining both statistical significance and $F$-distributions in its lower dimensional representation (Figure \ref{fig:eval_real}D, \ref{fig:f_corr_alga_all}). While SimCLR had an $F$-rank-ratio close to unity, comparable to other supervised ordinations – indicating high classification accuracy for clustering these data – its preservation scores for local and global structures were the lowest among all ordinations. Taken together, these multiple evaluations suggest that $F$-MDS is best suited for representing ecological data in terms of global ordering and group-based patterns, while its local structure representation is comparable to that of UMAP.

\subsection*{$F$-MDS adjusts group clusters to visualize statistical differences.} \label{sec:vis_fmds}
After evaluating $F$-MDS using different quality metrics, we applied the method to understand how sample groups were differentiated in 2D representations of compositional datasets. The compositional datasets described earlier were used for visualization, noting that their group differences were not distinguishable under 2D representations with metric MDS ($p_\mathbf{z}=0.645,\ 0.5$), whereas in the original dimension, they are statistically more distinct ($p_\mathbf{x}=0.004\ \text{and} <0.1$, respectively). To detect structural changes in these representations, visualizations were generated using metric MDS and $F$-MDS with varying hyperparameters, and the results were compared by calculating the inter-centroid distance and variance of each group.

\begin{figure}[h!]
    \centering
    \includegraphics[width=5.2in]{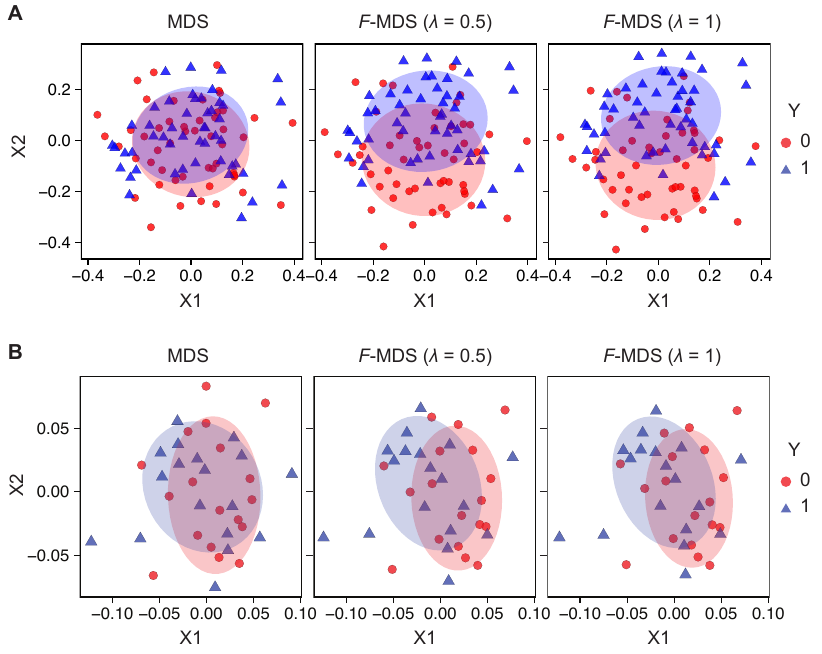}
    \caption{\textbf{\textit{F}-informed MDS can visualize group-based data discrimination.} Two-dimensional representations of (A) a simulated compositional dataset and (B) an algal microbiome dataset, generated using metric MDS and $F$-informed MDS ($F$-MDS) with two hyperparameter settings ($\lambda=$ 0.5, 1). Ellipses of respective colors are drawn with a confidence level of 0.68.}
    \label{fig:vis}
\end{figure}

For the simulated dataset, we observed a shift in group centroids when hyperparameter $\lambda$ increased from zero to one, with a unidirectional shift that separated the groups (Figure \ref{fig:vis}). The increase in inter-group centroid distances was more pronounced at lower hyperparameter values ($\lambda < 0.4$), with an increase rate of $0.28\pm 0.04\ (\lambda^{-1})$, compared to a slower rate at higher values ($0.07\pm 0.03\ \lambda^{-1},\ n=3$; \ref{fig:fmds_rep_analysis}A). The finding suggests that the re-arrangement of the MDS representation, constrained by the majorization, slowed down and stabilized at high $\lambda$. This is consistent with our earlier iteration number analysis (Figure \ref{fig:sim_iter}), where the iteration number reached a minimum, satisfying the objective $p_\mathbf{z}\approx p_\mathbf{x}$ as the hyperparameter was reached unity. Additionally, we observed a steady decrease in each group's variance by approximately $\sim\!12-14\%$ as $\lambda$ increased from 0 and 1 (\ref{fig:fmds_rep_analysis}B,C), as measured from the eigendecomposition of their covariances. This indicates that $F$-MDS further differentiated the groups by reducing their respective variances.

Similar trends were observed in the bacterial community dataset, where inter-group centroid distances were greater with non-zero hyperparameters than with metric MDS (Figure \ref{fig:vis}B). However, no statistical significance was observed in the correlation due to variability ($p = 0.521$, Spearman correlation, \ref{fig:fmds_rep_analysis}D). The primary (long-axis) variances of each 2D group representation also consistently decreased by 24\% and 18\% per $\lambda$ for each group ($y=0,\ 1$ respectively, \ref{fig:fmds_rep_analysis}E), while the decrease was less evident in the secondary (short-axis) variances (\ref{fig:fmds_rep_analysis}F). Further comparisons across the 2D representations with different $\lambda$ values showed that the sample distributions remained similar, suggesting that $F$-MDS still did not significantly alter the metric MDS configuration.

\begin{figure}[h!]
    \centering
    \includegraphics[width=5.2in]{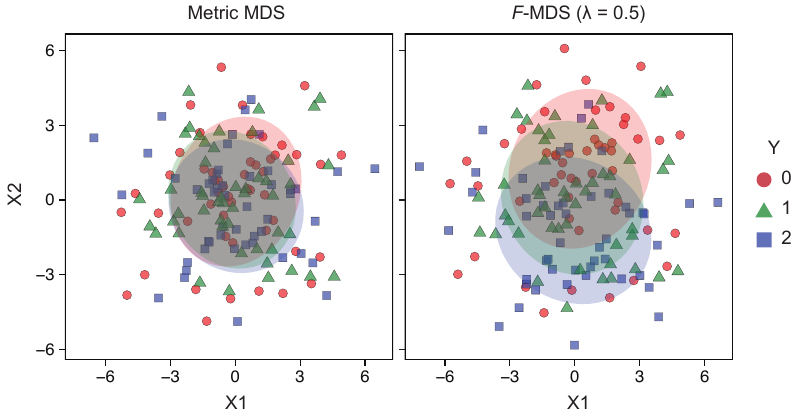}
    \caption{\textbf{Multi-class data is discriminated based on statistical inference.} Comparison of metric MDS and \textit{F}-informed MDS ($\lambda=0.5$) visualizations using a four-dimensional, three-class simulated dataset. Each dataset group follows normal distribution with the same covariance but different means (\ref{fig:sim_data_4d}). Ellipses of respective colors are drawn with a confidence level of 0.68.}
    \label{fig:vis_multiclass}
\end{figure}

\paragraph*{Differentiation of multi-group clusters.} To verify whether $F$-MDS can be applied to dataset consisting of more than two groups, we additionally considered a ternary dataset where each sample group followed a different four-dimensional normal distribution (Equation \ref{eq:ternary_distrib}, \ref{fig:sim_data_4d}). Similar to the earlier binary simulated dataset, the distribution was elliptically shaped, resulting in an indistinguishable representation in metric MDS ($p_\mathbf{z}=0.94$), whereas the original structures are statistically different ($p_\mathbf{x}<0.001$). Computing $F$-MDS with a hyperparameter $\lambda=0.5$ showed that the sample groups deviated from each other, making the differences more visible when ellipses with a 68\% confidence level were drawn (Figure \ref{fig:vis_multiclass}). While this deviation led to a statistically significant 2D representation ($p_\mathbf{z}=0.001$), the ternary class setting rendered the difference less pronounced compared to binary data representations.

\section*{Discussion}
In this work, we proposed a weakly supervised multidimensional scaling method based on the $F$-ratio and characterized its representations by varying hyperparameters and the iteration number of the majorization algorithm. Evaluations using simulated and bacterial community datasets showed that our $F$-informed MDS outperformed existing MDS-based and other dimension reduction methods in preserving local and global structures, as well as class-based information (e.g., statistical inference). The datasets used in this study represented cases where metric MDS failed to identify group differences through visualizations, partly due to the loss of such information during dimensionality reduction. This highlights the limitations of metric MDS in microbiome data analysis. Our dispersion-based approach expanded its applicability by introducing an additional constraint to adjust the MDS representation minimally. Since the adjusted representations significantly retain the sample distributions from metric MDS, this method remains applicable for visualizing microbial communities while enabling group-associated inferences directly from the visualizations.

One limitation of $F$-informed MDS, however, is that its computation relies on the initial configuration from metric MDS, and the computational cost increase with iterations of majorization and label permutations (a summary of method costs is provided in \ref{sec:app5}). Recent theoretical advances in  computational algorithms, including majorization \cite{Streeter23} or (stochastic) gradient descent \cite{Demaine21, zheng19} open doors for improving our framework toward unrestricted initialization and more efficient iterations. Additionally, empirical observations suggest that $F$-MDS representations became less reliant on the choice of hyperparameter at higher values. Once the relationship between the two objective-constituting terms is clarified, it could lead to a viable strategy for selecting an optimal hyperparameter without requiring recursive procedures such as cross-validation.

In summary, the proposed $F$-MDS provides a useful tool for visualizing high-throughput microbiome data while simultaneously delivering statistical testing results. We anticipate that this method will be beneficial for broader applications in microbiology and ecology data analysis. The R implementation of $F$-MDS is available at https://github.com/soob-kim/fmds.

\section*{Acknowledgments}
The authors acknowledge the MIT Office of Research Computing and Data for providing high performance computing resources that have contributed to the research results reported within this paper. We thank C. Belthangady and J.-Y. Lee for guidance in developing neural network models.

\bibliographystyle{abbrv}
\bibliography{refs}

\include{supp}
% \import{./}{supp.tex}

%%%%%%%%%%%%%%%%%%%%%%%%%%%%%%%%%%%%%%%%%%%%%%%%%%%%%%%%%%%%

\end{document}

%% file: supp.tex
\setlength{\parskip}{0.5em}

\makeatletter
\renewcommand\theequation{S\@arabic\c@equation}
\makeatother

\renewcommand\Affilfont{\normalsize}

%%%%%%%%%%%%%%%%%%%%%%%%%%%%%%%%%%%%%%%%%%%%%%%%%%%%%%%
%%%%%%%%%%%%%%%%%%%%%%%%%%%%%%%%%%%%%%%%%%%%%%%%%%%%%%%
% \begin{document}

\renewcommand*\contentsname{\normalsize List of Supplementary Notes}
\makeatletter
\renewcommand{\tableofcontents}{\@starttoc{toc}}
\renewcommand*\l@section{\@dottedtocline{1}{0em}{6.3em}}
\let\l@table\l@content
\makeatother

\makeatletter
\renewcommand{\listoffigures}{\@starttoc{lof}}
\renewcommand*\l@figure{\@dottedtocline{1}{0em}{3.5em}}
\let\l@table\l@figure
\makeatother

\makeatletter
\renewcommand{\listoftables}{\@starttoc{lot}}
\renewcommand*\l@table{\@dottedtocline{1}{0em}{4.5em}}
\let\l@table\l@table
\makeatother

\renewcommand\thesection{Supplementary Note S\arabic{section}}
\makeatletter
\renewcommand*{\@seccntformat}[1]{\csname the#1\endcsname\hspace{0.5em}}
\makeatother

\renewcommand{\figurename}{\!\!}
\let\standardthefigure\thefigure
\renewcommand\thefigure{Figure S\standardthefigure}
% \addtolength{\cftfignumwidth}{15pt}

\renewcommand{\tablename}{\!\!}
\let\standardthetable\thetable
\renewcommand\thetable{Table S\standardthetable}

\let\standardthealgorithm\thealgorithm
\renewcommand\thealgorithm{S\standardthealgorithm}

\setcounter{figure}{0}

\section{Mapping function for $F$-MDS.} \label{sec:app1}

In this section, we explain a procedure of deriving a mapping function $f_\mathbf{z}(F_\mathbf{x})$ that is used to minimize an objective function for computing $F$-informed MDS.
The confirmatory term of the objective function (Equation 5 of the main text) is specifically designed to minimize a difference in $p$-values testing for group difference by representations under the original $S$- and two-dimensions. 
These \textit{p}-values are obtained from an empirical distribution of the pseudo-$F$s by permuted labels \cite{anderson01}. 
In other words, by denoting a label set as $[y_i]$ ($i=1,\cdots N$) and its permuted set as $[y_i^\pi]$ with operator $\pi$, we express each $F$-statistic as
\begin{equation}
\begin{aligned}
F^\pi_\mathbf{x} &= 
\left(\frac{\sum_{i,j}d_{ij}^2}{2\sum_{i,j}d_{ij}^2\,\1\{y_i^\pi = y_j^\pi\}} -1 \right) \cdot (N-2), \\
F^\pi_\mathbf{z} &= 
\left(\frac{\sum_{i,j}\|\mathbf{z}_i - \mathbf{z}_j\|_2^2}{2\sum_{i,j}\|\mathbf{z}_i - \mathbf{z}_j\|_2^2\,\1\{y_i^\pi = y_j^\pi\}} -1 \right) \cdot (N-2).
\label{eq:map1}
\end{aligned}
\end{equation}
% \end{align} 

It should be noted that the pseudo $F$-ratios can range differently based on how the samples are projected onto 2D. To scale between these two ratios and minimize the difference in $p$-values, we consider a mapping $f_\mathbf{z}: F_\mathbf{x} \rightarrow F_\mathbf{z}$ from a pair of ratios $(F_\mathbf{x}^{\pi_1}, F_\mathbf{z}^{\pi_2})$ where each element was ordered from two independent sets of repeated permutations. Figure below exemplifies how the $F$-ratios at each dimensionality scaled differently and depended on the choice of hyperparameter $\lambda$. In both settings, the $F$-ratio in $S$-dimension ranged between zero and six whereas in 2D representation the ratio ranged in [0,4]. The hyperparameter, however, impacted the detailed relationship between ordered $F_\mathbf{x}$ and ordered $F_\mathbf{x}$, when compared to each other. 

\begin{figure}[h]
    \centering
    \includegraphics[width=5.5in]{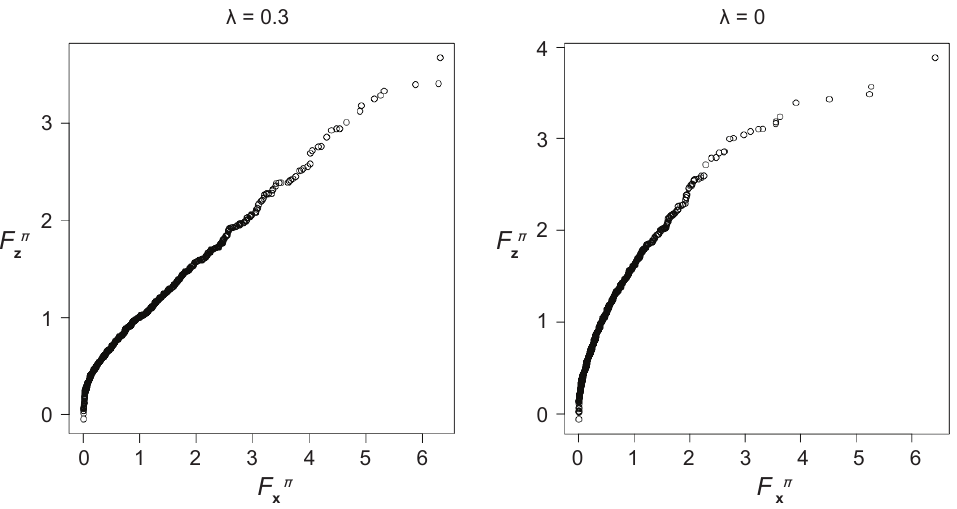}
    \caption*{Figure: Mapping pseudo-$F$'s between two dimensionalities. Each $F_\mathbf{x} \text{-} F_\mathbf{z}$ relationship was obtained by permuting labels over 1,000 times, and by setting a hyperparameter $\lambda$ that is used to perform the majorization.}
    \label{fig:mds-fval}
\end{figure}

Using the mapping function $f_\mathbf{z}$, we now derive an exact form of confirmatory term proposed in the $F$-MDS objective function (Equation 5 of main text). The proposed form $\left|F_\mathbf{z} - f_\mathbf{z}(F_\mathbf{x})\right|$ measures how the 2D representation accurately reflected the $S$-dimensional distance structure with respect to the sample labels. Having the difference close to zero indicates that no additional update with the representation is required (let us denote it as $\mathbf{Z}^*$). Now by substituting the confirmatory term $\left|F_\mathbf{z} - f_\mathbf{z}(F_\mathbf{x})\right|$ with Equation \ref{eq:map1}, we define $\1\{y_i^\pi = y_j^\pi\}\coloneqq\epsilon_{ij}$ and write
\begin{align}
\mathbf{Z}^* &= \argmin_\mathbf{Z}\,\left|F_\mathbf{z} - f_\mathbf{z}(F_\mathbf{x})\right| \\
& = \argmin_\mathbf{Z}\,\left|(N-2)\cdot\left(\frac{\sum_{i,j}\|\mathbf{z}_i - \mathbf{z}_j\|_2^2}{2\sum_{i,j}\epsilon_{ij}\|\mathbf{z}_i - \mathbf{z}_j\|_2^2} -1 \right) - f_\mathbf{z}(F_\mathbf{x})\right| \\
&= \argmin_\mathbf{Z}\,\left|\frac{\sum_{i,j}\|\mathbf{z}_i - \mathbf{z}_j\|_2^2}{2\sum_{i,j}\epsilon_{ij}\|\mathbf{z}_i - \mathbf{z}_j\|_2^2} -1 - \frac{f_\mathbf{z}(F_\mathbf{x})}{N-2} \right| \\
&= \argmin_\mathbf{Z}\,\left| \frac{\sum_{i,j}\|\mathbf{z}_i - \mathbf{z}_j\|_2^2 - 2\sum_{i,j}\epsilon_{ij}\|\mathbf{z}_i - \mathbf{z}_j\|_2^2 \cdot [1+f_\mathbf{z}(F_\mathbf{x})/(N-2)]}{2\sum_{i,j}\epsilon_{ij}\|\mathbf{z}_i - \mathbf{z}_j\|_2^2} \right| \label{eq:pre_approx}\\
&\approx \argmin_\mathbf{Z}\,\left| \sum_{i,j}\|\mathbf{z}_i - \mathbf{z}_j\|_2^2 - 2\sum_{i,j}\epsilon_{ij}\|\mathbf{z}_i - \mathbf{z}_j\|_2^2 \cdot \left(1+\frac{f_\mathbf{z}(F_\mathbf{x})}{N-2}\right) \right| \label{eq:approx}\\
&= \argmin_\mathbf{Z}\,\left|\sum_{i,j} \left[1- 2\epsilon_{ij} \left(1+\frac{f_\mathbf{z}(F_\mathbf{x})}{N-2}\right)\right] \|\mathbf{z}_i - \mathbf{z}_j\|_2^2\right|,
\end{align}
where Equation \ref{eq:approx} is approximated from Equation \ref{eq:pre_approx} to ensure that the same physical dimension as the stress term was obtained. By substituting expression in Equation \ref{eq:pre_approx} with the proposed confirmatory term, we conclude by deriving the exact form of the $F$-MDS objective as
\begin{equation}
    O_{\rm FMDS}(\mathbf{Z}) = \sum_{i,j} (d_{ij} - \| \mathbf{z}_i - \mathbf{z}_j \|_2 )^2 + \lambda\left|\sum_{i,j} \left[1- 2\epsilon_{ij} \left(1+\frac{f_\mathbf{z}(F_\mathbf{x})}{N-2}\right)\right] \|\mathbf{z}_i - \mathbf{z}_j\|_2^2\right|.
\label{eq:fmds_objective}
\end{equation}

\clearpage

\section{Majorization algorithm.} \label{sec:app2}

We seek a configuration $\mathbf{Z^*}= (\mathbf{z}_1^*, \cdots \mathbf{z}_N^*)\in \bbr^{N\times 2}$ that minimizes an objective term for $F$-MDS, $O_{\text{FMDS}}(\mathbf{Z})$ (Equation \ref{eq:fmds_objective}).
Previous work by Witten \& Tibshirani \cite{witten11} suggests that MDS-based ordinations can be computed by applying the Majorization (or Majorize-Minimization) algorithm where to a quadratic expression in terms of $\mathbf{Z}$ is analytically solved. In detail, for every $k=1,\cdots N$, a point $\mathbf{z}_k^*$ minimizing $O_{\text{FMDS}}(\mathbf{Z})$, while other configuration points fixed, is written as
\begingroup
\allowdisplaybreaks
\begin{align}
\mathbf{z}_k^* 
&= \argmin_{\mathbf{z}_k}O_{\rm FMDS}(\mathbf{Z} | \mathbf{z}_1,\cdots \mathbf{z}_{k-1}, \mathbf{z}_{k+1}, \mathbf{z}_N) \\
&= \argmin_{\mathbf{z}_k}\, \sum_{i,j} (d_{ij} - \|\mathbf{z}_i - \mathbf{z}_j\|_2)^2 + \lambda\left|\sum_{i,j} \left[1- 2\epsilon_{ij} \left(1+\frac{f_\mathbf{z}(F)}{N-2}\right)\right] \|\mathbf{z}_i - \mathbf{z}_j\|_2^2\right| \\
&= \argmin_{\mathbf{z}_k}\, \sum_{j=1}^N (d_{jk} - \|\mathbf{z}_j - \mathbf{z}_k\|_2)^2 + \lambda\delta(\mathbf{z})\sum_{j=1}^N \left[1- 2\epsilon_{jk} \left(1+\frac{f_\mathbf{z}(F)}{N-2}\right)\right] \cdot \|\mathbf{z}_j - \mathbf{z}_k\|_2^2 \\
&= \argmin_{\mathbf{z}_k}\, \sum_{j=1}^N \left[1+\lambda\delta(\mathbf{z}) \left(1- 2\epsilon_{jk} \left(1+\frac{f_\mathbf{z}(F)}{N-2}\right)\right) \right] 
\|\mathbf{z}_k-\mathbf{z}_j\|_2^2 
- 2d_{jk}\|\mathbf{z}_k-\mathbf{z}_j\|_2
\label{eq:appc_mm1}
\end{align}
\endgroup
where we have defined $\delta (\mathbf{z})$ as
\begin{align}
\delta (\mathbf{z}) = \text{sign}\left\{\,\sum_{i,j} \left[1- 2\epsilon_{ij} \left(1+\frac{f_\mathbf{z}(F)}{N-2}\right)\right] \|\mathbf{z}_i - \mathbf{z}_j\|_2^2\right\}.
\end{align}

As described by \cite{borg97b}, applying the algorithm starts with majorizing with Equation \ref{eq:appc_mm1},
\begin{equation}
\sum_{j=1}^N\,
\left[1+\lambda\delta(\mathbf{z}) \left(1- 2\epsilon_{jk} \left(1+\frac{f_\mathbf{z}(F)}{N-2}\right)\right) \right] \|\mathbf{z}_k-\mathbf{z}_j\|_2^2 -
2d_{jk}\,\frac{\sum_{s=1}^2 (z_{ks}-z_{js})(\Tilde{z}_{ks}-z_{js})}{\|\Tilde{\mathbf{z}}_k-\mathbf{z}_j\|_2}, \label{eq:appc_mm2}
\end{equation}
where $\Tilde{\mathbf{z}}_k$ is a fixed term (not updated) while ${\mathbf{z}}_k$ still remains as a variable. We further assume that a change of mapping function $f_\mathbf{z}(F)$ is negligible and that $\delta(\mathbf{z})$ remains constant during the iteration (e.g., a small change in $\mathbf{z}_k$ from metric MDS). These allow us to approximate Equation \ref{eq:appc_mm2} with a quadratic expression in terms of $\mathbf{z}$ and proceed to its minimization analytically.
To find its minimum at $\mathbf{z}_{k}=\mathbf{z}_{k}^\dagger$, a derivative is taken with respect to $z_{ks}$ and is set to zero.
In other words, we obtain
\begin{align}
\begin{split}
0 = \sum_{j=1}^N\, \left[1+\lambda\delta(\mathbf{z}) \left(1- 2\epsilon_{jk} \left(1+\frac{f_\mathbf{z}(F)}{N-2}\right)\right) \right] (z_{ks}^\dagger - z_{js}) 
- d_{jk}\frac{\Tilde{z}_{ks}-z_{js}}{\|\Tilde{\mathbf{z}}_k-\mathbf{z}_j\|_2}.
\label{eq:mm_balance}
\end{split}
\end{align}

Noting that for a balanced design where $\sum_{j=1}^N \epsilon_{jk}= {N}/{2}$, for $k=1,\cdots N$, we rewrite Equation \ref{eq:mm_balance} and analytically obtain $z_{ks}^\dagger$:
\begin{align}
\begin{split}
& \sum_{j=1}^N\, \left[1+\lambda\delta(\mathbf{z}) \left(1- 2\epsilon_{jk} \left(1+\frac{f_\mathbf{z}(F)}{N-2}\right)\right) \right] z_{ks}^\dagger 
= \left( N - \frac{N\lambda\delta(\mathbf{z})f_\mathbf{z}(F)}{N-2} \right) z_{ks}^\dagger \\
&\qquad= \sum_{j=1}^N\, \left[1+\lambda\delta(\mathbf{z}) \left(1- 2\epsilon_{jk} \left(1+\frac{f_\mathbf{z}(F)}{N-2}\right)\right) \right] z_{js} + d_{jk}\frac{\Tilde{z}_{ks}-z_{js}}{\|\Tilde{\mathbf{z}}_k-\mathbf{z}_j\|_2}.
\end{split}
\end{align}
\begin{align}
\begin{split}
\therefore z_{ks}^\dagger = 
&\frac{(N-2)}{N(N-2) - N\lambda\delta(\mathbf{z}) f_\mathbf{z}(F)} \\
&\quad\times\left\{\sum_{j=1}^N\, \left[1+\lambda\delta(\mathbf{z}) \left(1- 2\epsilon_{jk} \left(1+\frac{f_\mathbf{z}(F)}{N-2}\right)\right) \right]  z_{js} + d_{jk}\frac{\Tilde{z}_{ks}-z_{js}}{\|\Tilde{\mathbf{z}}_k-\mathbf{z}_j\|_2}\right\}.
\label{eq:update_element}
\end{split}
\end{align}
Finally, rewriting Equation \ref{eq:update_element} in a vector form gives us an update rule of $\mathbf{Z}$ as below:
\begin{align}
\begin{split}
\mathbf{z}_k \gets
\frac{(N-2)}{N(N-2) - N\lambda\delta(\mathbf{z}) f_\mathbf{z}(F)}\cdot \left\{\sum_{j=1}^N\, \left[1+\lambda\delta(\mathbf{z}) \left(1- 2\epsilon_{jk} \left(1+\frac{f_\mathbf{z}(F)}{N-2}\right)\right) \right]  \mathbf{z}_{j} + d_{jk}\frac{\mathbf{z}_k -\mathbf{z}_j}{\|{\mathbf{z}}_k-\mathbf{z}_j\|_2}\right\}.
\label{eq:mm_update}
\end{split}
\end{align}

\clearpage

\section{Human gut microbiome dataset.} \label{sec:app3}

We retrieved two sets of human gut microbiome data from a publicly available repository \cite{Pasolli16}, previously generated from Shotgun Metagenomic Sequencing, where the reads were merged at the genus level \cite{reiman20}. 
The first dataset contained a gut microbiome derived from 118 healthy and 114 liver cirrhosis patients from a single study \cite{Qin14}.
The second includes samples from a human gut of 217 healthy and 223 patients with type 2 diabetes (T2D) from two separate studies \cite{Qin12, Karlsson13}.

For both datasets, a phylogenetic tree was generated using phyloT \cite{Letunic23} where branch lengths were uniformly assigned with unity, resulting in 268 (cirrhosis) and 216 (T2D) features or taxa respectively. Taxonomy and abundance table were obtained using the phylogenetic tree and the merged reads, respectively, which were then integrated via \texttt{phyloseq} (Bioconductor v3.18). Pairwise distance matrix $\mathbf{D}$ was computed based on the weighted Unifrac \cite{lozupone07}.

\clearpage

\section{Neural network model and architecture.} \label{sec:app4}

We sought to compare our $F$-MDS with neural network models that are used for dimensionality reduction. To convert compositional microbial abundance into a matrix with its phylogenetic information, we implemented PopPhy-CNN \cite{reiman20} architecture. Each converted matrix reflected a phylogenetic tree structure by bacterial 16S rRNA amplicon (amplicon sequence variant or ASV) and its relative abundance which is normalized by cumulative sum scaling (CSS) \cite{paulson13}.
Thirty-six bacterial community samples were retrieved and re-analyzed from the previous work \cite{kim22}. The samples represent balanced design of diatom-associated community with and without presence of the host. Each compositional sample was converted to a 2D array sized $10 \times 42$.
The data was randomly split into training and validation sets (6 and 30 each) using the stratified K-Fold.

\begin{figure}[ht]
    \centering
    \includegraphics[width=6in]{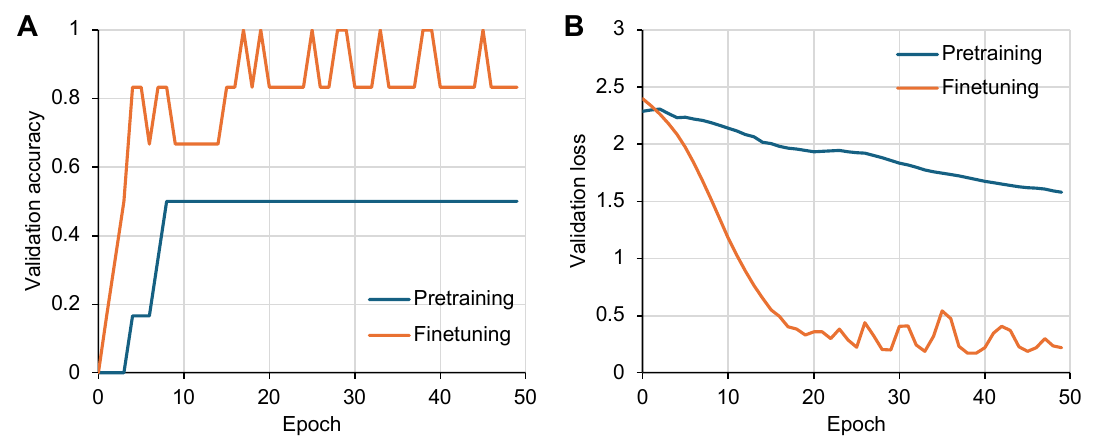}
    \caption*{Figure: Performance of self-supervised learning classifier by 50 training epochs. (A) Validation accuracy and (B) validation loss measured by categorical cross-entropy using bacterial community data represented as an image using PopPhy-CNN.}
    \label{fig:supp_nn_eval}
\end{figure}

The neural network architecture included an encoder consisting of one Gaussian noise filter, two 2D convolution layers (with kernel size of 5 by 3), and one fully connected layer with 32 output nodes. 
A self-supervised learning framework using SimCLR \cite{chen20} was chosen to explore its capability of arranging the microbiome data. The data augmentation was performed by applying random brightness and contrast filters with following parameters: (0.6, 0.2) for pretraining, (0.3, 0.1) for finetuning. In a pretraining step of SimCLR, a model was constructed by compiling encoder, projection head (two dense layers each of 32 output nodes), and one dense layer (10 output nodes). Site 1 and 2 datasets were individually trained and evaluated (30, 6 samples each) for the pretraining step, resulting a linear probing accuracy of 53.3\% after 50 training epochs (see Figure above). In the following finetuning step, the encoder is added with linear probe, resulting in a validation accuracy higher than 83.3\% after 50 epochs (see figure below). The trained encoder was used to obtain a 32 nodes-sized feature for each microbial community sample. The 32-dimensional feature was used for evaluating this neural network model with quality metrics.
Pairwise distance was calculated using $L_2$-squared metric to obtain Stress-1 and Shepard plot.

\clearpage

\section{Computational complexity.} \label{sec:app5}

We provide an upper bound of computational complexity of majorization algorithm for performing $F$-MDS.
Like most iteration-based optimizations, the computation cost of majorization was estimated on the basis of a single step.
We discuss the time complexity of each step outlined in Algorithms 1 and 2 of the main text.

For computing the mapping function (Algorithm 1), a pseudo-$F$-ratio is computed from a set of permuted labels $y^\pi$ and each of input matrices $d$, $\mathbf{z}$.
Each computation takes $\cO(N^2)$ operations with $N$ being the sample size.
The step repeats for a number of iteration, e.g., $p=999$, resulting in $\cO(2pN^2)$ operations.
Additional steps are taken to sort the lists of permuted $F$-ratios with $\cO(2N\log N)$.
In total, the complexity is $\cO(2pN^2 + 2N\log N)$.

For majorization step (Algorithm 2), the $F$-ratio is computed once and is mapped to $f_\mathbf{z}(F)$, taking $\cO(N^2 + \log N)$ operations.
Next, the sign of $F$-MDS confirmatory term $\delta(\mathbf{Z})$ is obtained and the step takes $\cO(N^2)$ operations.
Finally, the 2D representation $\mathbf{Z}$ is updated for every point, taking $\cO(N^2)$ operations.
Therefore, the complexity for one iteration of the majorization algorithm is $\cO(3N^2 + \log N)$.

In summary, the computational cost of performing $F$-MDS (unit iteration) is $\cO(2pN^2 + 3N^2 + 2N\log N + \log N) \approx \cO(pN^2)$. 
It is compared with other dimension reduction methods which is summarized below.

\begin{table}[h]
    \caption*{Comparison of time complexity between different dimensionality reduction methods.}
    \centering
    \begin{tabular}{c c c}
    \hline
       Method & Complexity & Algorithm \\
    \hline
        $F$-informed MDS\footnotemark & $\cO(pN^2)$ & Majorization \cite{borg97a} \\
        \multirow[c]{2}{*}{MDS} & $\cO(N^3)$ & Eigendecomposition \cite{Torgerson52} \\
         & $\cO(N\log N)$ & Divide-and-conquer \cite{Yang06} \\
        Supervised MDS$^1$ & $\cO(N^2)$ & Majorization \cite{witten11} \\
        UMAP & $\cO(N^{1.14})$ & NN-descent \cite{McInnes18, Dong11} \\
        \multirow{2}{*}{t-SNE$^1$} & $\cO(N^2)$ & Gradient descent \cite{maaten08} \\
         & $\cO(N\log N)$ & Barnes-Hut or dual-tree \cite{Maaten14} \\
        Isomap & $\cO(N^2\log N)$ & Dijkstra’s \cite{tenenbaum00, Pedregosa11} \\
    \hline
    \end{tabular}
    \label{tab:complexity}
\end{table}

\footnotetext[1]{Corresponds to a single iteration and does not represent a total complexity.}

\clearpage

\begin{figure}[h!]
    \centering
    \includegraphics[width=6.5in]{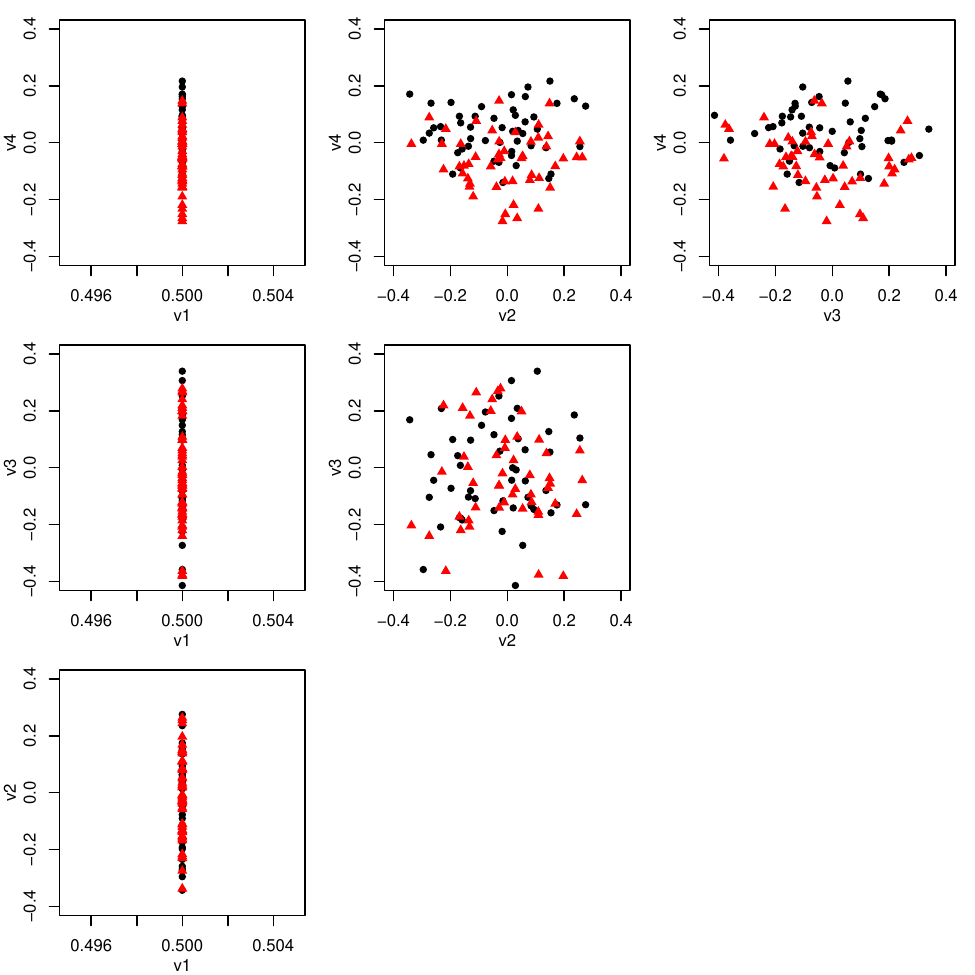}
    \caption[Pairs plot of simulated, binary dataset used in this study.]{Pairs plot of a simulated, binary dataset used in this study. Axes of each plot correspond to orthonormal vectors that were obtained through eigendecomposition of the design matrix $\mathbf{X}$. Alignment on the axis v1 indicates that the samples have been normalized (Equation 9, main text). Denoted with different colors and shapes are two groups that follow truncated normal distributions with different means but the same covariance (Equation 8, main text.) }
    \label{fig:sim_data_2d}
\end{figure}
\clearpage

\begin{figure}
    \centering
    \includegraphics[width=5in]{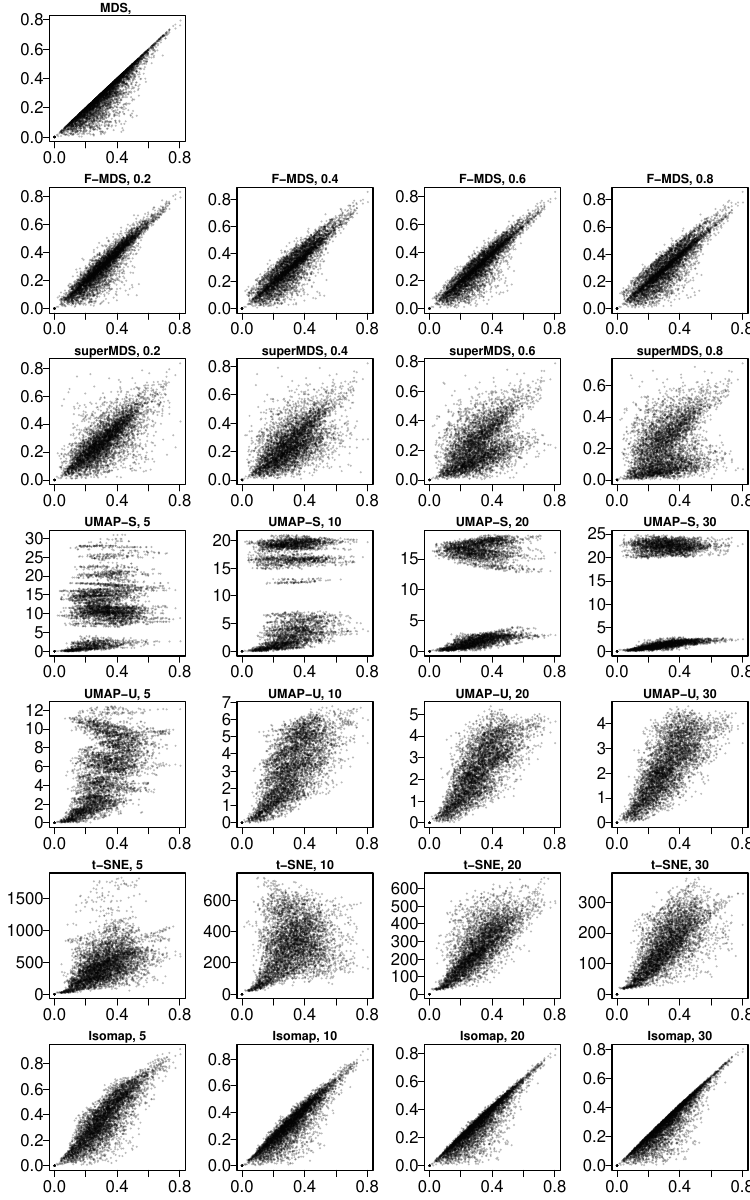}
    \caption[Shepard plot from 4D simulated dataset with seven ordination methods.]{Shepard plot of simulated dataset (first replicate) using seven dimension reduction methods. The plots are titled with the respective method and hyperparameter values as follows: $\lambda$, $F$-MDS; $\alpha$, superMDS; Nearest neighbors number, supervised (-S) or unsupervised (-U) UMAP; Perplexity, t-SNE; Shortest dissimilarities number, Isomap. X- and Y-axis denote the pairwise distance in the original and embedding dimensions, respectively.}
    \label{fig:shepard_sim_all}
\end{figure}
\clearpage

\begin{figure}
    \centering
    \includegraphics[width=5in]{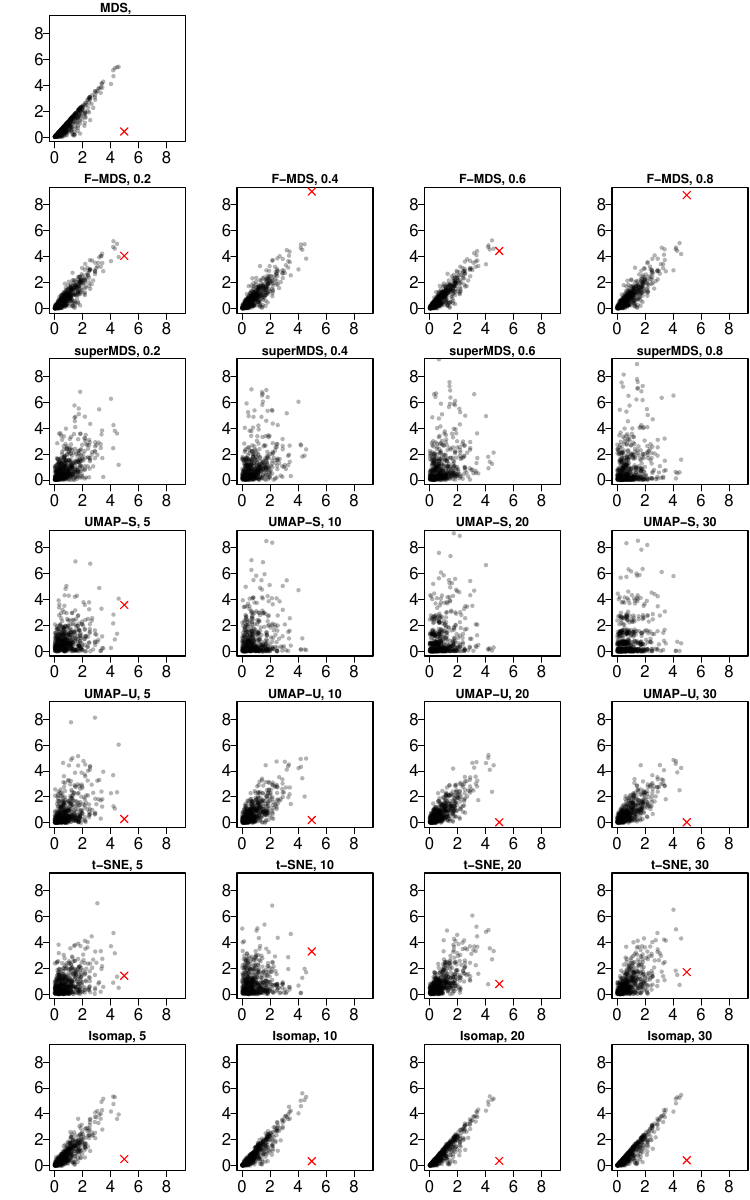}
    \caption[\textit{F}-correlation plot from simulated dataset.]{\textit{F}-correlation plot with simulated data. Pseudo $F$-ratios were calculated in the original dimension (x-axis) and in each dimension reduction method (y-axis) using first replicate of the simulated dataset. Pseudo $F$'s were calculated by randomly permuting labels by 500 times. Highlighted with red denotes the location of $F$'s from unpermuted labels. Each plot is titled with the method and hyperparameter used.}
    \label{fig:f_corr_sim_all}
\end{figure}
\clearpage

\begin{figure}
    \centering
    \includegraphics[width=5in]{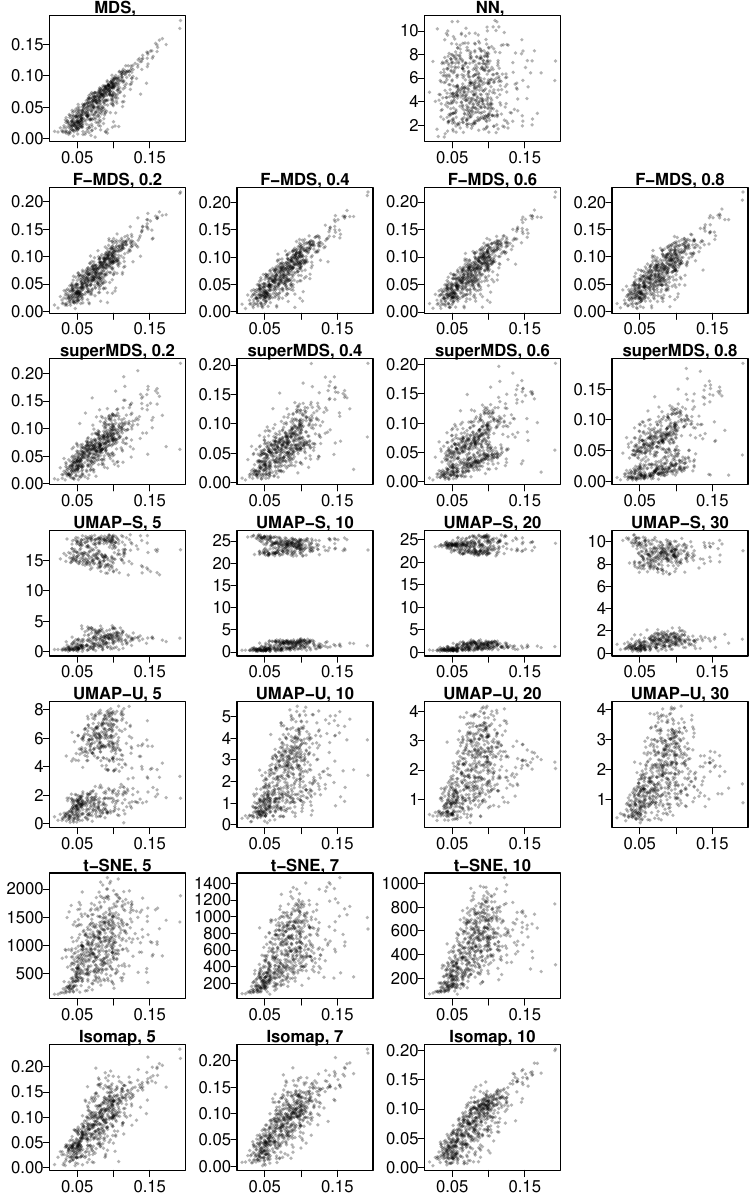}
    \caption[Shepard plot from algal microbiome dataset with eight ordination methods.]{Shepard plot of algal-associated bacterial community data using eight dimension reduction methods. The plots are titled with the respective method and hyperparameter values as follows: $\lambda$, $F$-MDS; $\alpha$, superMDS; Nearest neighbors number, supervised (-S) or unsupervised (-U) UMAP; Perplexity, t-SNE; Shortest dissimilarities number, Isomap; none, neural network (NN). X- and Y-axis denote distances in the original and embedding dimensions, respectively.}
    \label{fig:shepard_alga_all}
\end{figure}
\clearpage

\begin{figure}
    \centering
    \includegraphics[width=5in]{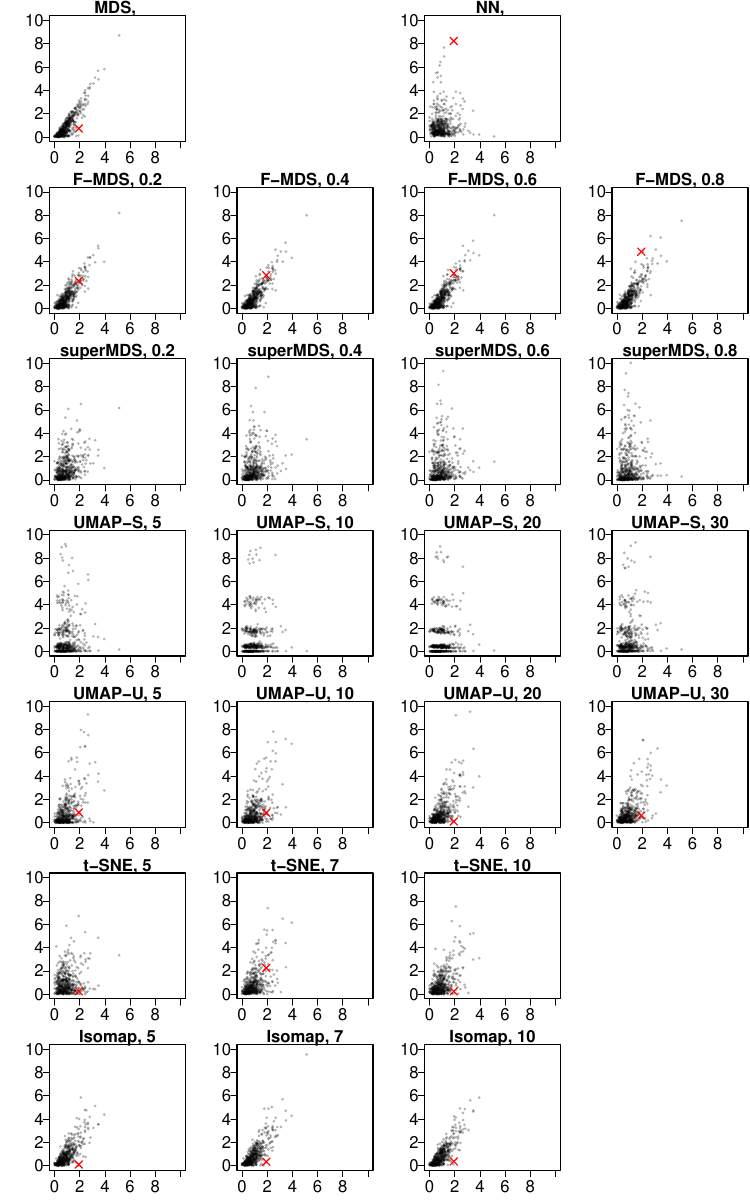}
    \caption[\textit{F}-correlation plot from algal microbiome dataset.]{\textit{F}-correlation plot from algal microbiome dataset. Pseudo $F$-ratios comparing the original dimension (x-axis) and from eight dimension reduction methods (y-axis) with algal microbiome data. Pseudo $F$'s were calculated by randomly permuting labels by 500 times. Highlighted with red denotes the location of $F$'s from unpermuted labels. Each plot is titled with the method and hyperparameter used.}
    \label{fig:f_corr_alga_all}
\end{figure}
\clearpage

\begin{figure}
    \centering
    \includegraphics[width=6.5in]{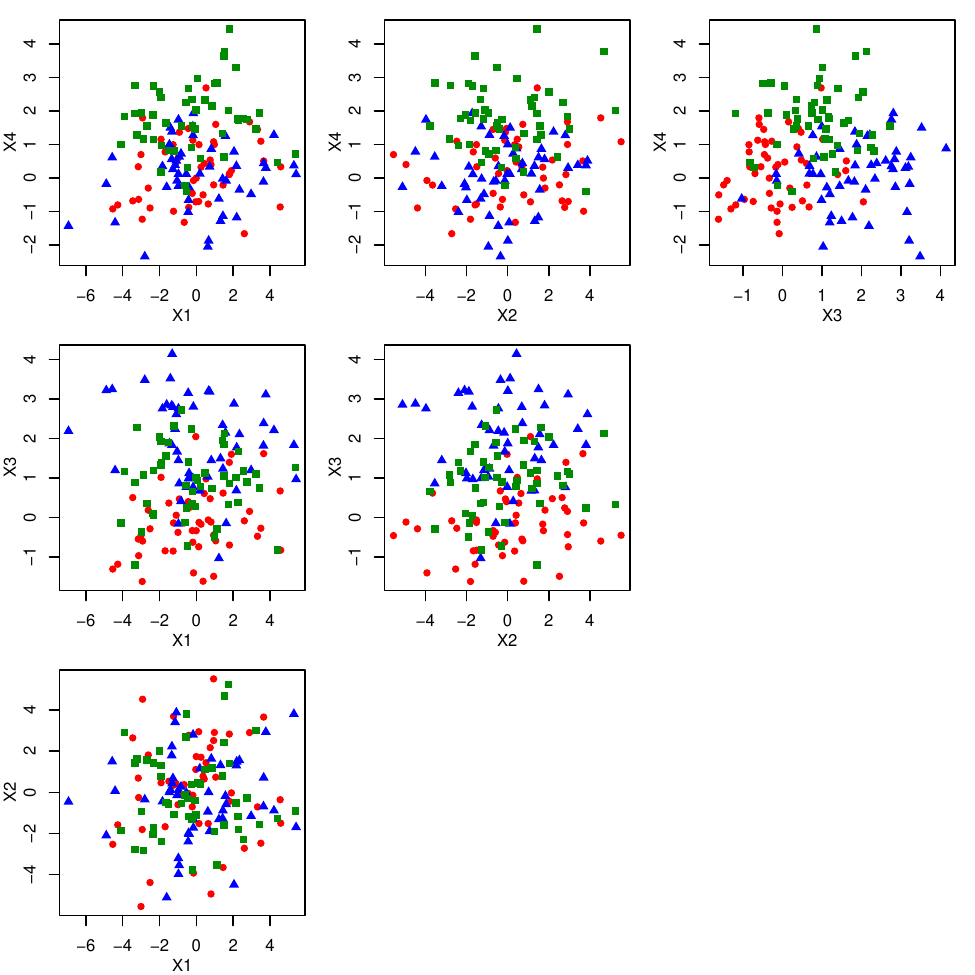}
    \caption[Pairs plot of simulated, trinary data.]{Pairs plot of trinary, four-dimensional simulated data. Each row and column corresponds to respective axis for the projection. Denoted with different colors or shapes are the trinary groups that follow normal distributions with different means but the same covariance (Equation 10, main text.) }
    \label{fig:sim_data_4d}
\end{figure}
\clearpage

\begin{figure}
    \centering
    \includegraphics[width=6.5in]{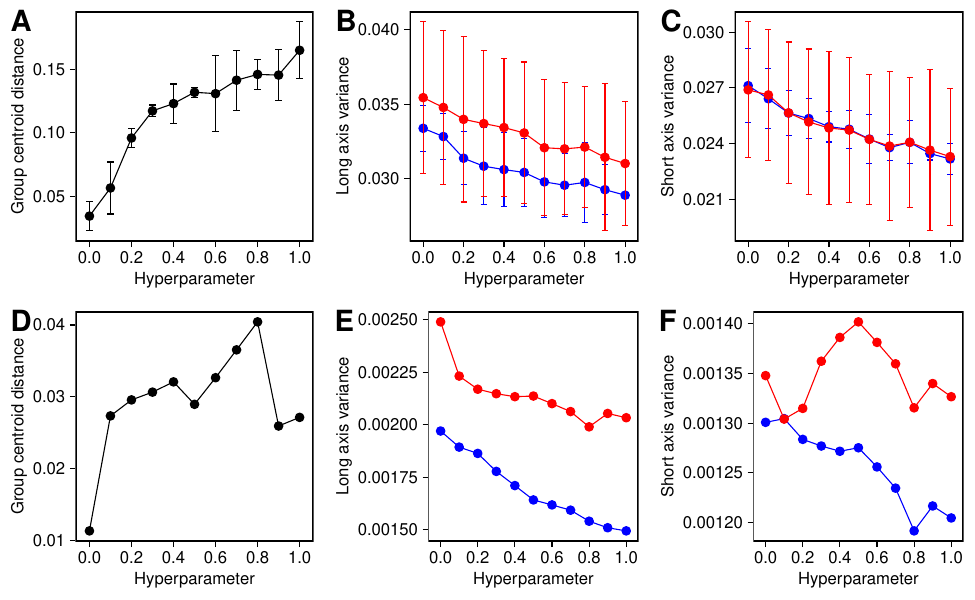}
    \caption[Cluster centroid and variances of $F$-MDS representations.]{Cluster centroid and variances of $F$-MDS representations for simulated dataset (A-C) and algal microbiome (D-F). A,D: Distance between group centroids. B,E: Variance of each group measured in its long-axis. C,F: Variance of each group measured in its short-axis. For simulated dataset error bars denote standard deviation of triplicates. For variance calculations, blue and red colors denote group 1 and 2, respectively.}
    \label{fig:fmds_rep_analysis}
\end{figure}
\clearpage

% \end{document}